%% file: JSAC_Revision_v2_2015_03_26.tex
\documentclass[journal,onecolumn,draftclsnofoot,12pt]{IEEEtran}

\usepackage{cite}
\usepackage{graphicx}
\usepackage{subfigure}
\usepackage{url}
\usepackage{stfloats}
\usepackage{epsfig}
\usepackage{amsmath}
\usepackage{amsfonts}
\usepackage{amssymb}
\usepackage{mathrsfs}
\usepackage[final,scrtime]{prelim2e}  

\usepackage{verbatim}
\usepackage{array}

\usepackage{algorithm}
\usepackage{algorithmic}

\usepackage{color}
\newcommand{\added}[1]{\textcolor{black}{#1}} 
\newcommand{\changed}[1]{\textcolor{black}{#1}} 
\newcommand{\addedv}[1]{\textcolor{black}{#1}} 
\newcommand{\changedv}[1]{\textcolor{black}{#1}} 

\newtheorem{thm}{Theorem}
\newtheorem{cor}{Corollary}
\newtheorem{lem}{Lemma}

\input{./Definitions.tex}

\newcommand{\bibdir}{../../Bibtex}

\begin{document}
%
\title{Heterogeneous Cellular Networks Using Wireless Backhaul: Fast Admission Control and Large System Analysis}
\author{Jian~Zhao,~\IEEEmembership{Member,~IEEE,}
        Tony Q. S. Quek,~\IEEEmembership{Senior Member,~IEEE,}
        and~Zhongding~Lei,~\IEEEmembership{Senior Member,~IEEE}%
\thanks{J. Zhao and Z. Lei are with the Institute for Infocomm Research, A*STAR, 1 Fusionopolis Way, \#21-01 Connexis, Singapore 138632 (e-mail: \{jzhao, leizd\}@i2r.a-star.edu.sg).}
\thanks{T. Q. S. Quek is with Singapore University of Technology and Design (e-mail: tonyquek@sutd.edu.sg). He is also with the Institute for Infocomm Research, A*STAR, Singapore.}
}%

\maketitle


\begin{abstract}
We consider a heterogeneous cellular network \changed{with densely underlaid small cell access points (SAPs)}. Wireless backhaul provides the data connection from the \changed{core network} to SAPs. To serve as many SAPs and their corresponding users as possible with guaranteed data rates, admission control of SAPs needs to be performed in wireless backhaul. Such a problem involves joint design of transmit beamformers, power control, and selection of SAPs. In order to tackle such a difficult problem, we apply $\ell_1$-relaxation and propose an iterative algorithm for the $\ell_1$-relaxed problem. The selection of SAPs is made based on the outputs of the iterative algorithm. This algorithm is fast and enjoys low complexity for small-to-medium sized systems. However, its solution depends on the actual channel state information, and resuming the algorithm for each new channel realization may be unrealistic for large systems. Therefore, we make use of random matrix theory and also propose an iterative algorithm for large systems. Such a large system iterative algorithm produces asymptotically optimum solution for the $\ell_1$-relaxed problem, which only requires large-scale channel coefficients irrespective of the actual channel realization. Near optimum results are achieved by our proposed algorithms in simulations.
\end{abstract}


\section{Introduction}
Fifth Generation (5G) mobile communication networks are expected to provide ubiquitous ultra-high data rate services and seamless user experience across the whole system~\cite{Andrews2014JSAC, 3GPPTR36913}, which makes it necessary to offload huge volumes of data and \addedv{large numbers of users} from macrocells to small cells. In 5G networks, hyper-dense deployment of small cells will be a key factor to achieving better spatial resource reuse and tremendous capacity enhancement to macrocells. \changed{The services of small cells are provided by small cell access points (SAPs), which are low power nodes only serving local-area users. In order to obtain the small cell user data, the SAPs are connected to the core network via backhaul.
}

As large numbers of small cells are underlaid with macrocells and more users are diverted \changed{from macrocells to small cells}, providing fast and reliable backhaul connection between \changed{the core network} and SAPs becomes a critical issue for such multi-tier heterogeneous networks~\cite{Chia2009Backhaul, Yang2014twc}. Wired backhaul, which uses copper or fiber cables, can provide high-rate data links between fixed stations. However, the cost to provide wired backhaul to all SAPs may be prohibitive when the number of SAPs is large. Moreover, certain locations that are difficult to be reached by wired access may restrict the universal deployment of wired backhaul. Wireless backhaul, which can overcome many of the drawbacks of wired backhaul, offers a cost-effective alternative~\cite{Hur2013mmWave, Lee2006wcnc, Bojic2013advanced, Ge2014FiveG}. Combined with renewable energy sources, SAPs with wireless backhaul can be established in a  self-sustained ``drop-and-play'' fashion, which is especially important in countries lacking reliable and ubiquitous power supply~\cite{Andrews2014JSAC}. Compared to wired backhaul, the management of wireless backhaul resources, e.g., power and spectrum, is more complicated due to finite power and radio spectrum constraints. 

Recent works have proposed analysis and design methods for backhaul technologies from many aspects.
A linear programming framework for determining optimum routing and scheduling of data flows in wireless mesh backhaul networks was proposed in~\cite{Viswanathan2006throughput}.
Zhao~\emph{et al.}~\cite{Zhao2012TWC} considered the problem of minimizing backhaul user data transfer in multicell joint processing networks, where algorithms involving joint design of transmit beamformers and user data allocation at base stations (BSs) were proposed to efficiently reduce the backhaul user data transfer.
Zhou~\emph{et al.}~\cite{zhou2013uplink} presented an information-theoretical study of an uplink multicell joint processing network in which the BSs are connected to a centralized processing server via rate-limited digital backhaul links employing the compress-and-forward technique. A similar scenario for the downlink of a cloud radio access network (Cloud-RAN) was investigated in~\cite{Park2013TSP}, where multivariate compression of different BSs' signals was exploited to combat additive quantization noise. 
The spectral efficiency and energy efficiency tradeoff in a homogeneous cellular network was investigated in~\cite{Rao2013tvt}, where the backhaul power consumption was taken into consideration. Considering the overall network power consumption including the backhaul, Shi~\emph{et al.}~\cite{Shi2014twc} proposed schemes to improve the energy efficiency in Cloud-RAN cellular networks. \added{For the Cloud-RAN backhauling, not only the routing of data but also the additional cost of baseband processing in the cloud infrastructure has been investigated in~\cite{Park2013TSP, Shi2014twc}.}
\added{Wireless backhaul technologies have been discussed in~\cite{Hur2013mmWave, Lee2006wcnc, Bojic2013advanced, Ge2014FiveG}. Hur~\emph{et al.}~\cite{Hur2013mmWave} proposed a beam alignment technique for millimeter wave wireless backhaul and investigated the tradeoff between array size and wind-induced movement. Lee~\emph{et al.}~\cite{Lee2006wcnc} provided several admission control schemes for multihop wireless backhaul network under rate and delay requirements. Flexible high-capacity hybrid wireless and optical mobile backhauling for small cells was investigated in~\cite{Bojic2013advanced}. The energy efficiencies of wireless backhaul networks for different system architectures and frequency bands were compared in~\cite{Ge2014FiveG}.
}

A vital task of wireless communications is to design schemes to meet the quality-of-service (QoS) requirements subject to given amount of resources. Resource management of wireless systems with QoS requirements has been discussed in~\cite{toelli2011decentralized, cai2012maxmin, zhai2014energy, Zhao2014TVT, ahmad2012coordinated}.
A decentralized method to minimize the sum transmit power of BSs under given QoS requirements was proposed in~\cite{toelli2011decentralized} for a multicell network, relying on limited backhaul information exchange between BSs.
Iterative algorithms to maximize the minimum QoS measure of users in multicell joint processing networks under per-BS power constraints were proposed in~\cite{cai2012maxmin}.
When it is not possible to meet the QoS requirements for all wireless stations with the given resources, only a subset of transmission links can be selected to be active.
Zhai \emph{et al.}~\cite{zhai2014energy} investigated the link activation problem in cognitive radio networks with single-antenna primary and secondary BSs and users. A price-driven spectrum access algorithm was proposed and the energy-infeasibility tradeoff was analyzed.
The uplink user admission control and user clustering under given QoS and power constraints were considered in~\cite{Zhao2014TVT}, where algorithms relying on the $\ell_1$-norm relaxation were proposed.
Transmission schemes using semidefinite relaxation and Gaussian randomization to select active antenna ports were proposed in~\cite{ahmad2012coordinated} to maximize the minimum user rate in multicell distributed antenna systems. 

In order to achieve enormous enhancement in spectral efficiency, the technique of ``massive multiple-input multiple-output (MIMO)'' is envisioned to be an important ingredient in 5G communication systems~\cite{marzetta2010noncooperative, rusek2013scaling}. In massive MIMO, the number of antennas equipped at BSs is much larger than that of active users in the same time-frequency channel.
Hoydis \emph{et al.}~\cite{hoydis2013massive} analyzed the number of required antennas in massive MIMO systems using different linear beamformers. They showed that more sophisticated beamformers may reduce the number of required antennas in massive MIMO systems to achieve the same performance.
Fernandes \emph{et al.}~\cite{fernandes2013inter} provided the asymptotic performance analysis of both the downlink and the uplink for a cellular network as the number of BS antennas tends to infinity.
Huang \emph{et al.}~\cite{Huang2013joint} studied joint beamformer design and power allocation in multicell massive MIMO networks, where efficient algorithms to maximize the minimum weighted QoS measure of all the users were proposed.

In this paper, we consider a heterogeneous cellular network with \changed{densely underlaid small cells}. The \changed{core network} are connected to the SAPs via wireless backhaul links, which provide the channel to transfer the small cell user data. A wireless backhaul hub (WBH) with multiple antennas is deployed to transmit wireless backhaul signals to the SAPs. Minimum data rate requirements must be satisfied when SAPs receive data from the WBH so that small cell users can be served at the required data rates. \changedv{In order to improve the utilization efficiency of wireless backhaul supported SAPs and let more SAPs and their corresponding small cell users be served, it is desirable for the WBH to support as many SAPs as possible.} Given QoS constraints at SAPs and power constraint at the WBH, we propose wireless backhaul transmission schemes aiming to admit the maximum number of SAPs into the network. Such transmission schemes involve joint design of beamformers and power control, as well as selection of subset of SAPs to be supported. We tackle this difficult problem by applying $\ell_1$-norm relaxation to the original non-convex non-smooth problem and utilize uplink-downlink duality to transform the transmit beamforming problem into an equivalent receive beamforming problem. Based on the optimality conditions, we propose an iterative algorithm that jointly update the values of the primal and dual variables. Such an algorithm is fast and we prove that it converges locally to the optimum solution. Based on the solution of the $\ell_1$-relaxed problem, we then iteratively remove the SAP that corresponds to the largest QoS gap until all the remaining SAPs can be supported.
Furthermore, we provide a large system analysis of the above SAP admission control problem. As the system dimensions become large, we show that certain system parameters may be approximated as deterministic quantities irrespective of the actual channel realization. Random matrix theory is leveraged to transform our proposed finite system iterative algorithm for large systems so that only large-scale statistical information of the channel is required. As long as the large-scale channel coefficients and the QoS requirements remain unchanged, the selected SAPs will satisfy the QoS and power constraints almost surely. 
Simulations are carried out to verify the proposed algorithms for finite and large systems. The proposed algorithms demonstrate fast convergence and low computational complexity. For the finite system iterative algorithm, the average number of admitted SAPs is very close to the optimum results. For the large system iterative algorithm, the results of the selected SAPs accurately match those obtained by Monte Carlo simulations. 

The remainder of the paper is organized as follows. In Section~\ref{sec:SystemModel}, we introduce the system model and formulate the SAP admission control problem. The iterative algorithm for finite systems and its convergence property are presented in Section~\ref{sec:MaxSAP}. The large system analysis and its corresponding iterative algorithm are presented in Section~\ref{sec:MassiveMIMO}. Section~\ref{sec:Simulation} shows the convergence behavior and the simulation results in a cellular scenario. Finally, conclusions are drawn in Section~\ref{sec:Conclusions}.

\textbf{Notation:}
we use bold uppercase letters to denote matrices and bold lowercase letters to denote vectors.
$\mathcal{CN}(0, \sigma^2)$ denotes a circularly symmetric complex normal zero mean random
variable with variance $\sigma^2$.
$(\cdot)^{T}$ and $(\cdot)^{H}$ stand for the transpose and conjugate transpose, respectively.
$\Complex^M$ and $\Real_{+}^M$ denote $M$-dimensional complex vectors and nonnegative real vectors, respectively.
$\EE{\cdot}$ stands for the mathematical expectation.
$\CB{\vec{u}_{i}}$ denotes the set made of $\vec{u}_{i}$, $\forall i$.
$\norm{\vec{x}}$, $\norm{\vec{x}}_0$ and $\norm{\vec{x}}_1$ stand for the Euclidean norm, the $\ell_0$-norm and the $\ell_1$-norm of the vector $\vec{x}$, respectively.
$\as$ denotes almost sure convergence.
$\rho(\mat{A})$ stands for the spectral radius of matrix $\mat{A}$.
$\vec{a}\circ\vec{b}$ denotes the Hadamard product of $\vec{a}$ and $\vec{b}$.
For a vector $\vec{q}$, $\vec{q}^{-1}$ stands for the element-wise inverse of $\vec{q}$.


\section{System Model and Problem Formulation}\label{sec:SystemModel}

\begin{figure}
  \centering
  \includegraphics[width=0.50\textwidth]{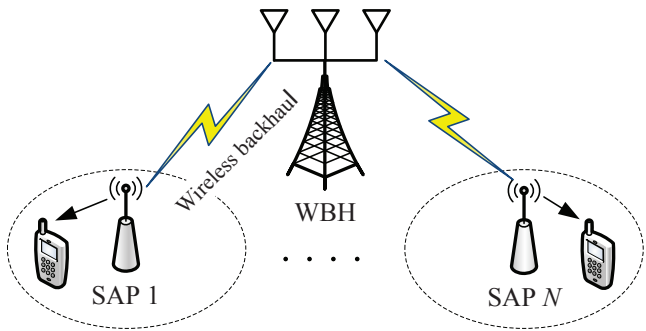}
  \caption{A heterogeneous cellular network with wireless backhaul.}
  \label{fig:Scenario}
\end{figure}

We consider a heterogeneous cellular network with $N$ underlaid single antenna SAPs as in Fig.~\ref{fig:Scenario}. The SAPs obtain their small cell user data from the core network via wireless backhaul. A WBH with $M$ antennas is responsible for transmitting wireless backhaul signals to the SAPs. The wireless backhaul spectrum is out-of-band, which does not interfere with users in the network. In order to meet the data rate requirements for serving small cell users, the received data at the SAPs must satisfy certain minimum rate requirements. In a heterogeneous cellular network, users that cannot be served by SAPs will have to be served by the macrocell base station (MBS) directly. \changedv{In order to improve the utilization efficiency of wireless backhaul supported SAPs and let more users be served by their corresponding small cells, it is desirable for the WBH to support as many SAPs as possible.} Under given QoS requirements at SAPs and transmit power constraint at WBH, we propose schemes aiming to admit the maximum subset of SAPs that can be simultaneously supported by the wireless backhaul.

We denote the wireless backhaul channel from the WBH to the $i$th SAP as $\vec{h}_{i}^H$, where $\vec{h}_{i} \in \Complex^M$, $\forall i = 1, \cdots, N$.
Linear transmit processing is applied at the WBH to deliver user data to the SAPs using wireless backhaul links. The transmit beamformer at the WBH for the $i$th SAP is denoted as $\vec{u}_{i} \in \Complex^M$, such that $\norm{\vec{u}_{i}} = 1$, $\forall i$. We denote the normalized data symbol for the $i$th SAP's users that is transmitted via wireless backhaul as $d_i$, where $\EE{\abs{d_i}^2} = 1$. The received signal at the $i$th SAP can be expressed as
\begin{equation}
    y_i = \vec{h}_i^H \vec{u}_i \sqrt{\frac{p_i}{M}} d_i + \sum_{j = 1, j \ne i}^{N} \vec{h}_i^H \vec{u}_j \sqrt{\frac{p_j}{M}} d_j + \eta_i
\end{equation}
where $\eta_i$ is the additive white Gaussian noise (AWGN) at the $i$th SAP, such that $\eta_i \isCN(0,n_i)$ and $n_i$ denotes the noise variance. The transmit power for data of the $i$th SAP is ${p_i}/{M}$. We denote the power constraint for backhaul transmission at the WBH as $P$. The backhaul transmit power must satisfy $\sum_{i=1}^N w_i {p_i}/{M} \le P$, where $w_i > 0$ is the weight for the $i$th transmit power.
The signal to interference-plus-noise ratio (SINR) at the $i$th SAP is given by
\begin{equation}\label{eq:SINR_D}
    \SINR_i^{\rm{D}} = \frac{\frac{p_i}{M} \abs{\vec{h}_i^H \vec{u}_i}^2}{\sum_{j =1, j \ne i}^{N} \frac{p_j}{M} \abs{\vec{h}_i^H \vec{u}_j}^2 + n_i}.
\end{equation}
In order to meet the QoS requirements of its small cell users, the receive SINR at each SAP must satisfy minimum SINR requirement. The SINR requirements at SAPs can be determined from the QoS requirements of their small cell users and they can be easily fed back to the core network\footnote{Denote the users of the $i$th small cell as $\mathbb{J}$. Assume the data rate requirement of the $i$th user to be $r_i$, $\forall i \in \mathbb{J}$. Then, the SINR requirement for the $i$th SAP must be chosen such that $\log_2 (1+\gamma_i) \ge \sum_{i \in \mathbb{J}} r_i$.}. We denote the SINR requirement for the $i$th SAP as $\gamma_i$, $\forall i$.







In heterogeneous networks, small cells are deployed within the macrocell to serve users so that the number of users needed to be served by the MBS can be reduced to minimum.
When the SAPs are densely deployed, the WBH may not be able to support all the SAPs simultaneously for given SINR and power constraints.
\changedv{The users that cannot be served by SAPs need to be served by the MBS. In order to let as many SAPs and their corresponding small cell users be served as possible, it is desirable to select the subset of SAPs with the maximum cardinality that can be simultaneously supported by the wireless backhaul. Since it is cheaper to build wireless backhaul compared to wired backhaul, we also minimize the total cost of building backhaul for small cells of the network in this way.} Such a problem of SAP admission control can be formulated as
\begin{equation}\label{opt:DL_L0}
\begin{array}{cl}
  \min\limits_{\vec{p}, \CB{\vec{u}_i}, \vec{x}} & \norm{\vec{x}}_0 \\
  \mbox{s.t.} & \displaystyle \SINR_i^{\rm{D}} \ge \frac{\gamma_i}{1+x_i}, \quad \forall i = 1, \cdots, N, \\
  & \frac{1}{M}\vec{w}^T \vec{p} \le P, \\
  & \vec{x} \ge \vec{0}
\end{array}
\end{equation}
where $\vec{w} = \SB{w_1, \cdots, w_N}^T$, $\vec{p} = \SB{p_1, \cdots, p_N}^T$, and $\vec{x} = \SB{x_1, \cdots, x_N}^T \in \Real_{+}^N$.
Here $x_i \ge 0$ represents the SINR gap of the $i$th SAP to satisfy its SINR requirement. If $x_i = 0$ in the solution, it shows that the SINR requirement $\gamma_i$ can be satisfied by the $i$th SAP. The objective function $\norm{\vec{x}}_0$ denotes the $\ell_0$-norm of $\vec{x}$, which is the number of non-zero elements in $\vec{x}$. 

\section{Analysis and Algorithm Design for Finite Systems}\label{sec:MaxSAP}
The problem \eqref{opt:DL_L0} is combinatorial and NP-hard due to the non-convex $\ell_0$-norm in the objective function~\cite{matskani2008convex}. Approximate solutions of non-convex optimization problems can be obtained by applying convex relaxation~\cite{Zhao2012TWC, Zhao2014TVT} and replacing the $\ell_0$-norm in~\eqref{opt:DL_L0} with its convex envelop, i.e., the $\ell_1$-norm. The $\ell_1$-relaxed problem can then be expressed as
\begin{equation}\label{opt:DL_L1}
\begin{array}{cl}
  \min\limits_{\vec{p}, \CB{\vec{u}_i}, \vec{x}} & \norm{\vec{x}}_1 \\ 
  \mbox{s.t.} & \text{constraints of }\eqref{opt:DL_L0}.
\end{array}
\end{equation}
The problem \eqref{opt:DL_L1} is still difficult to solve due to the need to jointly optimize the transmit power $\vec{p}$, transmit beamformers $\CB{\vec{u}_i}$, and the SINR gap $\vec{x}$. The following lemma shows that the downlink transmit optimization problem \eqref{opt:DL_L1} can be converted to an uplink receive optimization problem.






\begin{lem}\label{lem:UL_DL}
  The downlink transmit optimization problem \eqref{opt:DL_L1} is equivalent to the following dual uplink receive optimization problem
  \begin{equation}\label{opt:UL_L1}
  \begin{array}{cl}
    \min\limits_{\vec{q}, \CB{\vec{u}_i}, \vec{x}} & \sum_i x_i \\
    \mbox{s.t.} & \displaystyle \SINR_i^{\rm{U}} \ge \frac{\gamma_i}{1+x_i}, \quad \forall i = 1, \cdots, N, \\ 
    & \frac{1}{M}\vec{n}^T \vec{q} \le P, \\
    & \vec{x} \ge \vec{0}
  \end{array}
  \end{equation}
  where $\vec{n} = \SB{n_1, \cdots, n_N}^T$ and $\vec{q} = \SB{q_1, \cdots, q_N}^T \in \Real_{+}^N$. Here we define
\begin{equation}\label{eq:SINR_U}
  \SINR_i^{\rm{U}} = \frac{\frac{q_i}{M} \abs{\vec{u}_i^H \vec{h}_i}^2}{\sum_{j =1, j \ne i}^{N} \frac{q_j}{M} \abs{\vec{u}_i^H \vec{h}_j}^2 + w_i}.
\end{equation}
Furthermore, the optimum solution of $\CB{\vec{u}_i}$ and $\vec{x}$ in~\eqref{opt:DL_L1} are equal to those in~\eqref{opt:UL_L1}. The optimum solution of $\vec{p}$ in \eqref{opt:DL_L1} has one-to-one correspondence to the optimum solution of $\vec{q}$ in \eqref{opt:UL_L1}.
The problem~\eqref{opt:UL_L1} is a receive optimization problem in the dual uplink with the same SINR constraints, where $\vec{h}_i$ is the dual uplink channel from the $i$th SAP to the WBH and $w_i$ becomes the uplink noise variance at the $i$th SAP. Here $q_i/M$ represents the dual uplink transmit power of the $i$th SAP, and $\vec{u}_i^H$ is the uplink receive beamformer for the $i$th SAP transmission. 
\end{lem}

\begin{IEEEproof}
    See Appendix~\ref{apx:Proof_UL_DL}.
\end{IEEEproof}

We denote the optimum solution of \eqref{opt:DL_L1} as $\vec{p}^{\star}$, $\CB{\vec{u}_i^{\star}}$, $\vec{x}^{\star}$. According to Lemma~\ref{lem:UL_DL}, the optimum solution of \eqref{opt:UL_L1} can be denoted as $\vec{q}^{\star}$, $\CB{\vec{u}_i^{\star}}$, $\vec{x}^{\star}$. The following lemma shows that the optimum beamformer $\CB{\vec{u}_i^{\star}}$ can be determined from the optimum uplink power $\vec{q}^{\star}$.
\begin{lem}\label{lem:MMSE}
    The optimum receive beamformer $\CB{\vec{u}_i^{\star}}$ of \eqref{opt:UL_L1} is the minimum mean square error (MMSE) receiver, which can be obtained from the optimum uplink power $\vec{q}^{\star}$ as
    \begin{equation}\label{eq:mmse_BF}
        \vec{u}_i^{\star} = \vec{u}_i^{\rm{MMSE}}(\vec{q}^{\star}) = \frac{1}{\xi}\RB{\sum_{j =1, j \ne i}^N \frac{q_j^{\star}}{M} {\vec{h}_j \vec{h}_j^H}  + w_i \mat{I}}^{-1} \vec{h}_i, \quad \forall i = 1, \cdots, N
    \end{equation}
    where $\xi$ is a normalization factor such that $\norm{\vec{u}_i^{\star}} = 1$. The corresponding uplink SINR in \eqref{eq:SINR_U} is
    \begin{equation}\label{eq:sinr_UL}
        \SINR_i^{\rm{U}} \RB{\vec{q}^{\star}, \CB{\vec{u}_i^{\star}}} = \frac{q_i^{\star}}{M} \vec{h}_i^H \RB{\sum_{j = 1, j \ne i}^N \frac{q_j^{\star}}{M} {\vec{h}_j \vec{h}_j^H}  + w_i \mat{I}}^{-1} \vec{h}_i, \quad \forall i.
    \end{equation}
\end{lem}

\begin{IEEEproof}
    See Appendix~\ref{apx:Proof_MMSE}.
\end{IEEEproof}

For notational brevity, we define an equivalent channel matrix $\mat{G}$, where
\begin{equation}\label{eq:G}
    G_{ij} = \abs{\vec{h}_i^H \vec{u}_j}^2, \quad \forall i, j = 1, \cdots, N
\end{equation}
and we omit the dependence of $G_{ij}$ on $\vec{u}_j$.
After a change of variables $\qt_i = \log(\frac{q_i}{M})$, $\forall i$, the problem~\eqref{opt:UL_L1} can be equivalently expressed as
\begin{equation}\label{opt:GP}
\begin{array}{cl}
  \min\limits_{\vec{\qt}, \CB{\vec{u}_i}, \vec{x}} & \sum_i x_i \\
  \mbox{s.t.} & \displaystyle \log\frac{ \gamma_i \RB{\sum_{j=1, j \ne i}^N G_{ji}e^{\qt_j} + w_i} }{G_{ii}e^{\qt_i}} \le \log(1+x_i), \quad \forall i = 1, \cdots, N \\
  & \sum_{i=1}^N n_i e^{\qt_i} \le P, \\
  & \vec{x} \ge \vec{0}. 
\end{array}
\end{equation}
\changedv{For fixed $\CB{\vec{u}_i}$, the problem~\eqref{opt:GP} is a geometric programming (GP) problem with variables of $\vec{\qt}$ and $\vec{x}$. The standard way of solving a GP problem is using interior point algorithms, which is employed in software like cvx~\cite{cvx}. Considering $\CB{\vec{u}_i}$ also as optimization variables, one method to obtain solutions of~\eqref{opt:GP} is alternately optimizing $\CB{\vec{\qt}, \vec{x}}$ and $\CB{\vec{u}_i}$ by solving~\eqref{opt:GP} with fixed $\CB{\vec{u}_i}$ and updating the MMSE receiver $\CB{\vec{u}_i^{\rm{MMSE}}}$ using the obtained $\vec{\qt}$, respectively. However, such alternate optimization needs to solve~\eqref{opt:GP} using standard convex optimization software~\cite{cvx} in each iteration, which makes it relatively slow in practice. In the following, we provide a low complexity iterative algorithm.}



We associate the $i$th SINR constraint in \eqref{opt:GP} with the Lagrange dual variable $\nu_i$, the power constraint with $\mu$, and the nonnegativity constraint of $x_i$ with $\alpha_i$. The GP problem \eqref{opt:GP} satisfies Slater's condition. Hence, the  Karush-Kuhn-Tucker (KKT) conditions are necessary and sufficient for the optimality of \eqref{opt:GP}. The following lemma provides the optimality conditions for \eqref{opt:GP}, which is key to our iterative algorithm that solves \eqref{opt:DL_L1}.
\begin{lem}\label{thm1}
    The optimum primal and dual solutions of~\eqref{opt:GP} satisfy the following conditions
    \begin{IEEEeqnarray}{rCl}
      x_i^{\star}  &=& \max\RB{\nu_i^{\star} - 1, 0}, \quad \forall i = 1,\cdots, N; \label{eq:KKT1_} \\
      \frac{M \nu_i^{\star}}{q_i^{\star}} &=& \sum_{j =1, j \ne i}^N \frac{M G_{ij} \gamma_j \nu_j^{\star}}{(1+x_j^{\star}) G_{jj}q_j^{\star}} + \mu^{\star} n_i, \quad \forall i; \label{eq:KKT5_}\\
      \frac{(1+x_i^{\star}) G_{ii}q_i^{\star}}{M \gamma_i} &=& \sum_{j = 1, j \ne i}^N \frac{G_{ji}q_j^{\star}}{M} + w_i, \quad \forall i; \label{eq:KKT3_} \\
      \sum_{i=1}^N \frac{n_i q_i^{\star}}{M} &=& P; \label{eq:KKT4_} \\
      \mu^{\star} &>& 0; \quad \nu_i^{\star} > 0, \ \forall i.
    \end{IEEEeqnarray}
    The optimum downlink power $\vec{p}^{\star}$ of \eqref{opt:DL_L1} can be obtained directly from the optimum primal and dual solutions of~\eqref{opt:GP} as
    \begin{equation}\label{eq:p_opt}
        \frac{p_i^{\star}}{M} = \frac{M \gamma_i\nu_i^{\star}}{(1+x_i^{\star}) G_{ii} q_i^{\star} \mu^{\star}}, \quad \forall i.
    \end{equation}
    Furthermore, we have
    \begin{IEEEeqnarray}{rCl}
      \frac{(1+x_i^{\star}) G_{ii}p_i^{\star}}{M \gamma_i} &=& \sum_{j=1, j \ne i}^N \frac{G_{ij}p_j^{\star}}{M} + n_i, \quad \forall i;  \label{eq:KKT7_} \\
      \sum_{i=1}^N \frac{w_i p_i^{\star}}{M} &=& P. \label{eq:p_sum}
    \end{IEEEeqnarray}
\end{lem}

\begin{IEEEproof}
    See Appendix~\ref{apx:Proof_KKT}.
\end{IEEEproof}

\begin{cor}\label{cor1}
    The optimum downlink transmit power $\vec{p}^{\star}$ of \eqref{opt:DL_L1} and the optimum primal and dual solutions of \eqref{opt:GP} satisfy
\begin{IEEEeqnarray}{rCl}
  \nu_i^{\star} &=& \RB{\sum_{j=1, j \ne i}^N \frac{G_{ij} p_j^{\star}}{M} + n_i} \frac{\mu^{\star} q_i^{\star}}{M}, \quad \forall i = 1, \cdots, N;  \label{eq:nu_}\\
  \mu^{\star} &=& \sum_{i=1}^N \frac{M \gamma_i \nu_i^{\star} w_i}{(1+x_i^{\star}) G_{ii} q_i^{\star} P}; \label{eq:lambda_}\\
  \frac{\gamma_i}{(1+x_i^{\star})} &=& \SINR_i^{\rm{U}} \RB{\vec{q}^{\star}, \CB{\vec{u}_i^{\star}}} = \frac{q_i^{\star}}{M} \vec{h}_i^H \RB{\sum_{j=1, j \ne i}^N \frac{q_j^{\star}}{M} {\vec{h}_j \vec{h}_j^H}  + w_i \mat{I}}^{-1} \vec{h}_i, \quad \forall i.    \label{eq:sinr_UL_}
\end{IEEEeqnarray}
\end{cor}
Here \eqref{eq:nu_} is obtained by substituting \eqref{eq:p_opt} into \eqref{eq:KKT5_}. Eq.~\eqref{eq:lambda_} is obtained by substituting \eqref{eq:p_opt} into \eqref{eq:p_sum}. Eq.~\eqref{eq:sinr_UL_} is obtained by \eqref{eq:KKT3_} and Lemma~\ref{lem:MMSE}.

\begin{algorithm}
  \caption{Finite system iterative algorithm to solve \eqref{opt:DL_L1}}\label{alg:Alg1}
  \begin{algorithmic}[1]
    \STATE Initialization: $\nu_i \ge 1$ and $q_i > 0$, $\forall i$, such that $\sum_{i=1}^N n_i q_i = M P$.
    \REPEAT
    \STATE Store old values of $\vec{q}$
    \begin{equation}\label{step:q_old}
        \vec{\tilde{q}} = \vec{q}.
    \end{equation}
    \STATE Update
    \begin{equation}\label{step:q}
        \bar{q}_i = \frac{M \gamma_i}{\max(\nu_i,1) \cdot \RB{\vec{h}_i^H \RB{\sum_{j = 1, j \ne i}^N \frac{q_j}{M} {\vec{h}_j \vec{h}_j^H}  + w_i \mat{I}}^{-1} \vec{h}_i}}, \quad \forall i = 1, \cdots, N.
    \end{equation}
    \STATE Normalize
    \begin{equation}\label{step:q_norm}
        \vec{\bar{q}} = \frac{M P}{\vec{n}^T \vec{\bar{q}}} \vec{\bar{q}}.  
    \end{equation}
    \STATE Set
    \begin{equation}\label{step:Aitken}
      \vec{q} = \frac{1}{2}(\vec{\bar{q}} + \vec{\tilde{q}}).
    \end{equation}
    \STATE Calculate $\vec{u}_i$
    \begin{equation}\label{step:uj}
       \vec{u}_i = \frac{1}{\xi}\RB{\sum_{j = 1, j \ne i}^N \frac{q_j^{\star}}{M} {\vec{h}_j \vec{h}_j^H}  + w_i \mat{I}}^{-1} \vec{h}_i, \quad \forall i.
    \end{equation}
    \STATE Calculate the equivalent channel
    \begin{equation}\label{step:Gij}
      G_{ij} = \abs{\vec{h}_i^H \vec{u}_j}^2, \quad \forall i, j = 1, \cdots, N.
    \end{equation}
    \STATE Calculate $\mu$
    \begin{equation}\label{step:lambda}
        \mu = \sum_{i=1}^N \frac{M \gamma_i \nu_i w_i}{\max(\nu_i,1) \cdot G_{ii} q_i P}.
    \end{equation}
    \STATE Calculate the downlink power
    \begin{equation}\label{step:p}
        \frac{p_i}{M} = \frac{M \gamma_i \nu_i}{\max(\nu_i,1) \cdot G_{ii} q_i \mu}, \quad \forall i.
    \end{equation}
    \STATE Update $\nu_i$
    \begin{equation}\label{step:nu}
        \nu_i = \RB{\sum_{j = 1, j \ne i}^N \frac{G_{ij}p_j}{M} + n_i} \frac{\mu q_i}{M}, \quad \forall i.
    \end{equation}
    \UNTIL{$\abs{q_i - \tilde{q}_i} \le \epsilon$, $\forall i$.}
    \STATE Set $x_i = \max(\nu_i-1, 0)$, $\forall i$.
    \RETURN $\vec{x}$, $\vec{p}$ and $\CB{\vec{u}_i}$
  \end{algorithmic}
\end{algorithm}

Based on Lemma~\ref{thm1} and Corollary~\ref{cor1}, we propose Algorithm~\ref{alg:Alg1} that iteratively updates the values of the primal and dual variables to obtain the optimum $\vec{p}^{\star}$, $\CB{\vec{u}_i^{\star}}$ and $\vec{x}^{\star}$ of \eqref{opt:DL_L1}. In Algorithm~\ref{alg:Alg1}, we first store the old values of $\vec{q}$ in \eqref{step:q_old}. The fixed-point iteration~\eqref{step:q} and normalization~\eqref{step:q_norm} are obtained according to \eqref{eq:sinr_UL_} and \eqref{eq:KKT4_}, respectively. Here \eqref{step:Aitken} is to ensure contraction mapping for the algorithm, which will be made clear in the proof of Theorem~\ref{thm2}. After obtaining the uplink power $\vec{q}$, we calculate the corresponding MMSE receiver and the equivalent channel in \eqref{step:uj} and \eqref{step:Gij} according to \eqref{eq:mmse_BF} and \eqref{eq:G}, respectively. The value of $\mu$ in \eqref{step:lambda} is obtained according to \eqref{eq:lambda_}. The corresponding downlink power \eqref{step:p} is calculated according to \eqref{eq:p_opt} after getting the value of $\mu$, and the value of $\nu_i$ in \eqref{step:nu} is updated according to \eqref{eq:nu_}. In this algorithm, the optimum solution of $\vec{p}^{\star}$ is obtained directly in \eqref{step:p}. There is no need to perform the uplink-downlink power mapping~\cite{bengtsson2001optimal}. The convergence property of Algorithm~\ref{alg:Alg1} is shown in the following theorem.

\begin{thm}\label{thm2}
    Starting from an initial point that is sufficiently close to the optimum solution of \eqref{opt:UL_L1}, Algorithm~\ref{alg:Alg1} converges to the optimum solution of \eqref{opt:UL_L1} that satisfies the KKT conditions. 
\end{thm}

\begin{IEEEproof}
    See Appendix~\ref{apx:Proof_Converge}
\end{IEEEproof}

Theorem~\ref{thm2} shows that Algorithm~\ref{alg:Alg1} converges locally to the optimum solution of~\eqref{opt:DL_L1}. However, its range of convergence cannot be obtained from Theorem~\ref{thm2}, and the result may depend on the initialization point. Whether this algorithm can converge globally is still an open problem and is left for future work. In our simulations, we observe that its range of convergence is large and even random initialization will converge to correct results. 

\subsection{Connections With Max-Min SINR}
If all the SAPs can be supported, $\vec{x}^{\star} = \vec{0}$ in the solution of \eqref{opt:GP}. The values of $\nu_i$, $\forall i$, decrease monotonically towards zero in Algorithm~\ref{alg:Alg1}. In this case, the updates of power in \eqref{step:q}--\eqref{step:q_norm} become the power iteration steps in the max-min SINR algorithm of~\cite{cai2011unified}. Algorithm~\ref{alg:Alg1} will still converge. The output $\vec{p}$ and $\CB{\vec{u}_i}$ of Algorithm~\ref{alg:Alg1} are the power allocation and beamformers that maximize the minimum SINR of all SAPs in the system.

\subsection{Iterative SAP Removal}\label{subsec: IterativeRemoval}
Due to the convex relaxation, the solution of~\eqref{opt:DL_L1} may not be always optimal for the $\ell_0$-norm optimization problem~\eqref{opt:DL_L0}. Therefore, \eqref{opt:DL_L1} cannot be simply used as the substitution of~\eqref{opt:DL_L0} and we need to refine the selection of users based on the solution of~\eqref{opt:DL_L1}.
Since the solution of $x_i$ in~\eqref{opt:DL_L1} represents the gap of the $i$th SAP's SINR to satisfy its SINR requirement, it is natural for us to select those SAPs with small $x_i$. We propose to iteratively remove SAPs with decreasing values of $x_i$ until the remaining set of SAPs becomes feasible to satisfy the SINR and power constraints, i.e., until $x_i = 0$, $\forall i$. In this way, we obtain the final results for SAP admission control. 

\subsection{Admission Control Between SAPs and Users}\label{subsec: UserAdmission}

\addedv{The data rate requirements at the admitted SAPs can be guaranteed by the wireless backhaul with the proposed SAP admission control algorithm. After the admitted SAPs received data from the core network, they need to transmit those data to their corresponding small cell users. Existing  coordinated multipoint (CoMP) transmission schemes can be applied~\cite{toelli2011decentralized, cai2012maxmin}. However, due to the inter-SAP interference, there exists possibility that the users of the admitted small cells cannot be simultaneously supported with their given SINR requirements. If this happens, we need to perform user admission control within the admitted small cells in order to let as many users be served by their corresponding SAPs as possible. We briefly discuss the user admission control as follows. For simplicity of discussion, we assume each SAP serves one user. Multiple users per SAP case can be extended in a straightforward manner.}








\addedv{We denote the set of admitted small cells as $\mc{S}$ and the power constraint for the $i$th admitted SAP as $P_i$, where $i \in \mc{S}$. The channel gain between the $j$th SAP and the $i$th user is denoted as $g_{ij}$, where $i,j \in \mc{S}$. Within the set of admitted small cells, the user admission control problem can be formulated similar to~\eqref{opt:DL_L0}. After $\ell_1$-relaxation, we need to solve the following GP problem
\begin{equation}\label{opt:User}
\begin{array}{cl}
  \min\limits_{\vec{p}, \vec{x}} & \sum_i x_i \\
  \mbox{s.t.} & \displaystyle \frac{g_{ii}{p_i}}{ \sum_{j \in \mc{S}, j \ne i} g_{ij}{p_j} + n_i } \ge \frac{\gamma_i}{1+x_i}, \quad \forall i \in \mc{S} \\
  & p_i \le P_i, \quad \forall i \in \mc{S} \\
  & \vec{x} \ge \vec{0},
\end{array}
\end{equation}
where $p_i$ and $x_i$ denote the transmit power for the $i$th SAP and the SINR gap for the $i$th user, respectively. The problem~\eqref{opt:User} is similar to~\eqref{opt:GP}, with the only difference in the power constraints. The problem~\eqref{opt:User} with per-SAP power constraints can be solved by solving a series of weighted sum power constrained problems employing Algorithm~\ref{alg:Alg1}. Following similar discussion as \cite{zhang2012multi, dahrouj2010coordinated}, it can be shown that the optimum values of~\eqref{opt:User} is equal to the optimum values of the problem $\max_{\CB{w_i \ge 0}} f(\CB{w_i})$, where $f(\CB{w_i})$ with $\CB{w_i}$ as the parameter denotes the optimum objective value of the following weighted sum power constrained problem
\begin{equation}\label{opt:User_SumP}
\begin{array}{rcl}
  f(\CB{w_i}) = & \min\limits_{\vec{p}, \vec{x}} & \sum_i x_i \\
  & \mbox{s.t.} & \displaystyle \frac{g_{ii}{p_i}}{ \sum_{j=1, j \ne i}^N g_{ij}{p_j} + n_i } \ge \frac{\gamma_i}{1+x_i}, \quad \forall i \in \mc{S} \\
  & & \sum_{i \in \mc{S}} w_i p_i \le \sum_{i \in \mc{S}} w_i P_i,  \\
  & & \vec{x} \ge \vec{0}.
\end{array}
\end{equation}
where $w_i \ge 0$, $\forall i \in \mc{S}$, are the weights of power and also the parameters of $f(\CB{w_i})$. The problem~\eqref{opt:User_SumP} can be solved by the proposed Algorithm~\ref{alg:Alg1}. Furthermore, it can be shown that $P_i - p^{*}_i$ is a subgradient for $w_i$, where $p^{*}_i$ is the solution of~\eqref{opt:User_SumP} for the current iteration $\CB{w_i}$. Therefore, the solution of~\eqref{opt:User} can be obtained by the projected subgradient method: for a set of given weights $\CB{w^{(n)}_i}$, we can obtain $f \RB{\CB{w_i^{(n)}}}$ and the corresponding $\CB{p^{*}_i}$ by the proposed Algorithm~\ref{alg:Alg1} in the $n$th iteration; after that, the weights can be updated as
\begin{equation}
  w^{(n+1)}_i = \max\RB{w^{(n)}_i + t_n(P_i - p^{*}_i), 0}, \quad \forall i \in \mc{S}.
\end{equation}
By iteratively updating the weights $\CB{w_i}$ and solving~\eqref{opt:User_SumP}, we can obtain the solution of~\eqref{opt:User}. Finally, iterative removal of users with their corresponding SAPs as discussed in Section~\ref{subsec: IterativeRemoval} can be applied according to the solution of~\eqref{opt:User} until all the remaining users can be admitted.}

\section{Asymptotic Analysis and Algorithm Design for Large Systems}\label{sec:MassiveMIMO}
The algorithm proposed in Section~\ref{sec:MaxSAP} requires instantaneous channel state information. However, the instantaneous channel state information may change rapidly with time, which can incur frequent resumptions of Algorithm~\ref{alg:Alg1} to determine the SAP admission. On the other hand, 5G wireless networks are envisioned to be characterized by large numbers of antennas at WBH and dense deployments of SAPs. Under those circumstances, certain system parameters tend to become deterministic quantities that only depend on large-scale channel statistical information and the QoS requirements at SAPs. The large-scale channel statistical information include pathloss, shadowing and antenna gain, which do not change rapidly with time. The optimum power allocation $\vec{p}^{\star}$ in \eqref{opt:DL_L1} and the final selection of SAPs based on $\vec{x}^{\star}$ will also tend to be deterministic irrespective of the actual channel changes. In the following, we use random matrix theory to provide an iterative algorithm for the SAP admission control problem of large systems. Such an iterative algorithm solves \eqref{opt:DL_L1} in the asymptotic optimum sense. As long as the large-scale channel coefficients and the QoS requirements remain unchanged, the selected SAPs using large system analysis will almost surely be the same as the results obtained by the method of Section~\ref{sec:MaxSAP} based on instantaneous channel state information. 

We assume the number of transmit antennas $M$ at WBH and the number of SAPs $N$ go to infinity while the ratio $N/M$ remains bounded, i.e., let $M, N \rightarrow \infty$ while $0 < \lim \inf \frac{N}{M} \le \lim \sup \frac{N}{M} < \infty$. Such an assumption is abbreviated as $M \rightarrow \infty$. We use $\as$ to denote almost sure convergence, where $f(\vec{x}) \as a$ means $a$ is the deterministic equivalent of $f(\vec{x})$ as $M \rightarrow \infty$.
The following fading channel model is used for large system analysis
\begin{equation}\label{eq:h_RMT}
  \vec{h}_i = \sqrt{d_i} \vec{\tilde{h}}_i, \quad \forall i = 1, \cdots, N
\end{equation}
where $d_i$ represents the large-scale fading coefficient between the WBH and the $i$th SAP. The $\vec{\tilde{h}}_i$ denotes a normalized channel vector whose elements are independent and identically distributed (i.i.d.) $\CN(0,1)$ random variables. 




Under the channel model~\eqref{eq:h_RMT}, the uplink SINR~\eqref{eq:sinr_UL} using MMSE receive beamformers can be expressed as
\begin{IEEEeqnarray}{rCl}
  \SINR_i^{\rm{U}} \RB{\vec{q}, \CB{\vec{u}_i^{\rm{MMSE}}}} &=& \frac{q_i}{M} \vec{h}_i^H \RB{\sum_{j = 1, j \ne i}^N \frac{q_j}{M} {\vec{h}_j \vec{h}_j^H}  + w_i \mat{I}}^{-1} \vec{h}_i \\
  &=& \frac{q_i d_i}{M} \vec{\tilde{h}}_i^H \RB{\sum_{j = 1, j \ne i}^N \frac{q_j d_j}{M} {\vec{\tilde{h}}_j \vec{\tilde{h}}_j^H}  + w_i \mat{I}}^{-1} \vec{\tilde{h}}_i \quad \forall i. \label{eq:SINR_UL_RMT}
\end{IEEEeqnarray}
Calculating the uplink SINR~\eqref{eq:SINR_UL_RMT} involves matrix inversion of dimension $M$, which becomes increasingly complex as $M \rightarrow \infty$. However, the following lemma shows that the uplink SINR~\eqref{eq:SINR_UL_RMT} will tend to deterministic quantities asymptotically. Such deterministic quantities only depend on large-scale fading coefficients $\CB{d_i}$ irrespective of the instantaneous channel state information $\CB{\vec{h}_i}$.



\begin{lem}\label{lem:UL_SINR_RMT}
As $M \rightarrow \infty$, the uplink SINRs~\eqref{eq:SINR_UL_RMT} using MMSE receive beamformers approach deterministic quantities for given $\vec{q}$, i.e.,
\begin{equation}
  \SINR_i^{\rm{U}} \RB{\vec{q}, \CB{\vec{u}_i^{\rm{MMSE}}}} \as q_i d_i \varphi_i(\vec{q}), \quad \text{as } M \rightarrow \infty
\end{equation}
where
\begin{equation}\label{eq:phi_i}
  \varphi_i (\vec{q}) = \RB{w_i + \frac{1}{M}\sum_{j = 1, j \ne i}^N \frac{q_j d_j}{1+q_j d_j \varphi_i (\vec{q})}}^{-1}, \quad \forall i = 1, \cdots, N.
\end{equation}
\end{lem}
\begin{IEEEproof}
     By applying the trace lemma~\cite{bai2009spectral, couillet2011random} and the right-sided correlation model~\cite{silverstein1995empirical, couillet2011deterministic}, the proof follows~\cite[Remark 6.1]{couillet2011random}. 
\end{IEEEproof}
The value of $\varphi_i (\vec{q})$ is defined implicitly by fixed-point equation~\eqref{eq:phi_i} and it is deterministic for given $\vec{q}$. 



Similar to the uplink SINR using MMSE receive beamformers, the equivalent channels of the system using MMSE beamformers also approach deterministic quantities asymptotically. Using the channel model~\eqref{eq:h_RMT}, we have
\begin{equation}\label{eq:Gij_RMT}
  \frac{1}{M} \abs{\vec{h}_i^H \vec{u}_j^{\rm{MMSE}}}^2 = \frac{d_i}{M} \abs{\vec{\tilde{h}}_i^H \vec{u}_j^{\rm{MMSE}}}^2, \quad \forall i, j = 1, \cdots, N.
\end{equation}
Depending on whether $i = j$, the equivalent channels approach different deterministic values asymptotically. The following lemma shows the asymptotic equivalent channels when $i = j$.
\begin{lem}\label{lem:Gjj_RMT}
As $M \rightarrow \infty$, the equivalent channels $\frac{1}{M}\abs{\vec{h}_i^H \vec{u}_i^{\rm{MMSE}}}^2$ using MMSE beamformers approach deterministic quantities, i.e.,
\begin{equation}
  \frac{1}{M}\abs{\vec{h}_i^H \vec{u}_i^{\rm{MMSE}}}^2 \as \frac{d_i \varphi_i^2 (\vec{q})}{- \varphi'_i (\vec{q})}
\end{equation}
where
\begin{equation}
  \varphi'_i (\vec{q}) = {-\varphi_i (\vec{q})}\cdot \RB{w_i + \frac{1}{M}\sum_{j = 1, j \ne i}^N \frac{q_j d_j}{\RB{1+q_j d_j \varphi_i (\vec{q})}^2}}^{-1}.
\end{equation}
\end{lem}
\begin{IEEEproof}
     By utilizing the right-sided correlation model~\cite{silverstein1995empirical, couillet2011deterministic} and its derivatives, the proof follows~\cite[Theorem 2]{Wagner2012LargeSystem}. 
\end{IEEEproof}

The asymptotic equivalent channel when $i \ne j$ can be obtained by the following lemma.
\begin{lem}\label{lem:Gjl_RMT}
As $M \rightarrow \infty$, the equivalent channels $\abs{\vec{h}_i^H \vec{u}_j^{\rm{MMSE}}}^2$ using MMSE beamformers approach deterministic quantities, i.e.,
\begin{equation}
  \abs{\vec{h}_i^H \vec{u}_j^{\rm{MMSE}}}^2 \as \frac{d_i}{(1+q_i d_i \varphi_j(\vec{q}))^2}, \quad \text{when } i \ne j. 
\end{equation}
\end{lem}
\begin{IEEEproof}
     By utilizing Lemma~\ref{lem:Gjj_RMT} and applying the matrix inversion lemma~\cite{silverstein1995empirical} and the rank-1 perturbation lemma~\cite{bai2007signal}, the proof follows~\cite[Theorem 2]{Wagner2012LargeSystem}. 
\end{IEEEproof}

\begin{algorithm}
  \caption{Large system iterative algorithm to solve \eqref{opt:DL_L1}}\label{alg:Alg2}
  \begin{algorithmic}[1]
    \STATE Initialization: $\varphi_i (\vec{q}) >0$, $\nu_i > 0$ and $q_i > 0$, $\forall i$, such that $\sum_{i=1}^N n_i q_i = M P$.
    \REPEAT
    \STATE Store old values of $\vec{q}$
    \[\label{step:q_old_RMT}
        \vec{\tilde{q}} = \vec{q}.
    \]
    \STATE Update
    \begin{equation}
        \bar{q}_i = \frac{\gamma_i}{\max(\nu_i,1) \cdot \varphi_i (\vec{q}) d_i}, \quad \forall i = 1, \cdots, N.
    \end{equation}
    \STATE Normalize
    \begin{equation}
        \vec{\bar{q}} = \frac{M P}{\vec{n}^T \vec{\bar{q}}} \vec{\bar{q}}.  
    \end{equation}
    \STATE Set
    \begin{equation}\label{step:Aitken_RMT}
      \vec{q} = \frac{1}{2}(\vec{\bar{q}} + \vec{\tilde{q}}).
    \end{equation}
    \STATE Update $\varphi_i (\vec{q})$
    \begin{equation}
      \varphi_i (\vec{q}) = \frac{1}{w_i + \frac{1}{M}\sum_{j = 1, j \ne i}^N \frac{q_j d_j}{1+q_j d_j \varphi_i (\vec{q})}}, \quad \forall i.
    \end{equation}
    \STATE Calculate the corresponding $\varphi'_i (\vec{q})$
    \begin{equation}
      \varphi'_i (\vec{q}) = \frac{-\varphi_i (\vec{q})}{w_i + \frac{1}{M}\sum_{j = 1, j \ne i}^N \frac{q_j d_j}{\RB{1+q_j d_j \varphi_i (\vec{q})}^2}}, \quad \forall i.
    \end{equation}
    \STATE Calculate $\mu$
    \begin{equation}\label{step:lambda_RMT}
        \mu = \sum_{i=1}^N \frac{- M \nu_i \gamma_i \varphi'_i (\vec{q})}{\max(\nu_i,1) \cdot q_i d_i \varphi_i^2 (\vec{q})}.
    \end{equation}
    \STATE Calculate the downlink power
    \begin{equation}
        p_i = \frac{- M \nu_i \gamma_i \varphi'_i (\vec{q})}{\max(\nu_i,1) \cdot q_i d_i \varphi_i^2 (\vec{q}) \mu}, \quad \forall i.
    \end{equation}
    \STATE Update $\nu_i$
    \begin{equation}
        \nu_i = \RB{n_i + \frac{1}{M}\sum_{j=1, j \ne i}^N \frac{p_j d_i}{\RB{1+q_i d_i \varphi_j(\vec{q})}^2}} \frac{\mu q_i}{M}, \quad \forall i.
    \end{equation}
    \UNTIL{$\abs{q_i - \tilde{q}_i} \le \epsilon$, $\forall i$.}
    \STATE Set $x_i = \max(\nu_i-1, 0)$, $\forall i$. 
    \RETURN $\vec{x}$, $\vec{p}$ and $\vec{q}$
  \end{algorithmic}
\end{algorithm}

As $M \rightarrow \infty$, we can substitute the deterministic equivalents of the system parameters into Algorithm~\ref{alg:Alg1} and obtain an iterative algorithm for large systems that solves~\eqref{opt:DL_L1} as shown in Algorithm~\ref{alg:Alg2}. Lemma~\ref{lem:UL_SINR_RMT} is applied to \eqref{step:q} when updating $\vec{q}$. Lemma~\ref{lem:Gjj_RMT} is applied to \eqref{step:lambda} and \eqref{step:p} when updating $\nu$ and $\vec{p}$. Lemma~\ref{lem:Gjl_RMT} is applied to \eqref{step:nu} when updating $\CB{\nu_i}$. Algorithm~\ref{alg:Alg2} obtains the asymptotically optimum solutions for $\vec{p}^{\star}$, $\vec{q}^{\star}$ and $\vec{x}^{\star}$ of~\eqref{opt:DL_L1} using MMSE beamformers when $M \rightarrow \infty$. Note the beamformers $\CB{\vec{u}_i^{\star}}$ do not approach deterministic quantities because their values depend on the actual channel realization. In real system implementations, the outputs of Algorithm~\ref{alg:Alg2} can be calculated and stored. Depending on the actual channel state information $\CB{\vec{h}_i}$, the MMSE beamformers $\CB{\vec{u}_i^{\star}}$ can be directly obtained using \eqref{eq:mmse_BF}. \added{Unlike Algorithm~\ref{alg:Alg1}, there is no need to resume the algorithm for different channel realizations. In addition, the large-scale fading coefficients of SAPs changes slowly in reality. Algorithm~\ref{alg:Alg2} can be re-invoked as soon as the QoS requirements of SAPs change. Therefore, Algorithm~\ref{alg:Alg2} can save lots of computation in large systems. After we obtain the output of Algorithm~\ref{alg:Alg2}, iterative SAP removal can be applied based on the result of $\vec{x}$ achieved by Algorithm~\ref{alg:Alg2}.}




\subsection{Complexity Analysis}
\added{For the finite system analysis, the iterative Algorithm~\ref{alg:Alg1} needs to perform iterative updates of the primal and dual variables. The most computational intensive steps in the iterations are the steps involving matrix inversion and matrix-vector multiplication in~\eqref{step:q} and \eqref{step:uj}, where each step has the complexity of $O(N M^2+M^3)$ for each user. Considering all the $N$ users, the complexity of each iteration is $O(N^2 M^2 + N M^3)$. By applying the convergence results of fixed-point algorithms~\cite{feyzmahdavian2012contractive}, we obtain that the complexity of Algorithm~\ref{alg:Alg1} is $O( (N^2 M^2 + N M^3)\log c_1^{-1} )$, where $c_1$ is a constant that determines the convergence speed of iterative operations in Algorithm~\ref{alg:Alg1}. Following similar discussions, we obtain that the computational complexity of the large system iterative Algorithm~\ref{alg:Alg2} is $O( N^2 \log c_2^{-1} )$, where $c_2$ is a constant that determines the convergence speed of iterative operations in Algorithm~\ref{alg:Alg2}. If we use the exhaustive search method, the computational complexity will be $O( 2^N (N^2 M^2 + N M^3) N_{\rm{max}} )$, where $N_{\rm{max}}$ denotes the maximum number of iterations using the max-min SINR algorithm~\cite{cai2011unified} to check the feasibility of the selected SAPs. Therefore, the proposed algorithms enjoy polynomial complexity with respect to $N$, which is much lower than the exponential complexity required by the exhaustive search method.
}


\section{Simulation Results}\label{sec:Simulation}

\subsection{Convergence Behavior of Finite System Iterative Algorithm}
To verify the iterative algorithm for finite systems, we consider a heterogeneous cellular network, where the number of antennas at the WBH is $M = 3$ and the number of SAPs is $N = 4$. Each SAP has a single receive antenna for the wireless backhaul.  The SINR requirement at each SAP for the wireless backhaul is set at $\gamma_i = 3.01$dB, $\forall i = 1, \cdots, N$. The noise variance at each SAP is normalized to $n_i = 1$ Watt, $\forall i$. The transmit power constraint for wireless backhaul at the WBH is $P = 10$ Watt. In the whole simulations, the weights for the transmit power are set to $w_i = 1$, $\forall i$. A randomly generated channel is used to show the convergence result. We compare the results obtained by two methods: Algorithm~\ref{alg:Alg1} and \emph{alternate updates using cvx} (AUC). In AUC, we solve~\eqref{opt:DL_L1} by alternatively solving~\eqref{opt:GP} with fixed $\CB{\vec{u}_i}$ \addedv{using cvx~\cite{cvx}} to obtain uplink power $\vec{q}$ and calculating the MMSE receiver $\CB{\vec{u}_i^{\rm{MMSE}}}$ using the obtained $\vec{q}$. In the end, we map the obtained uplink power $\vec{q}$ into the corresponding downlink power $\vec{p}$ that achieves the same SINR using uplink-downlink power mapping~\cite{schubert04solution}.



\begin{figure*}
\centering
    \subfigure[Convergence result of power $p_i$, $\forall i$]
    {\includegraphics[width=0.48\textwidth]{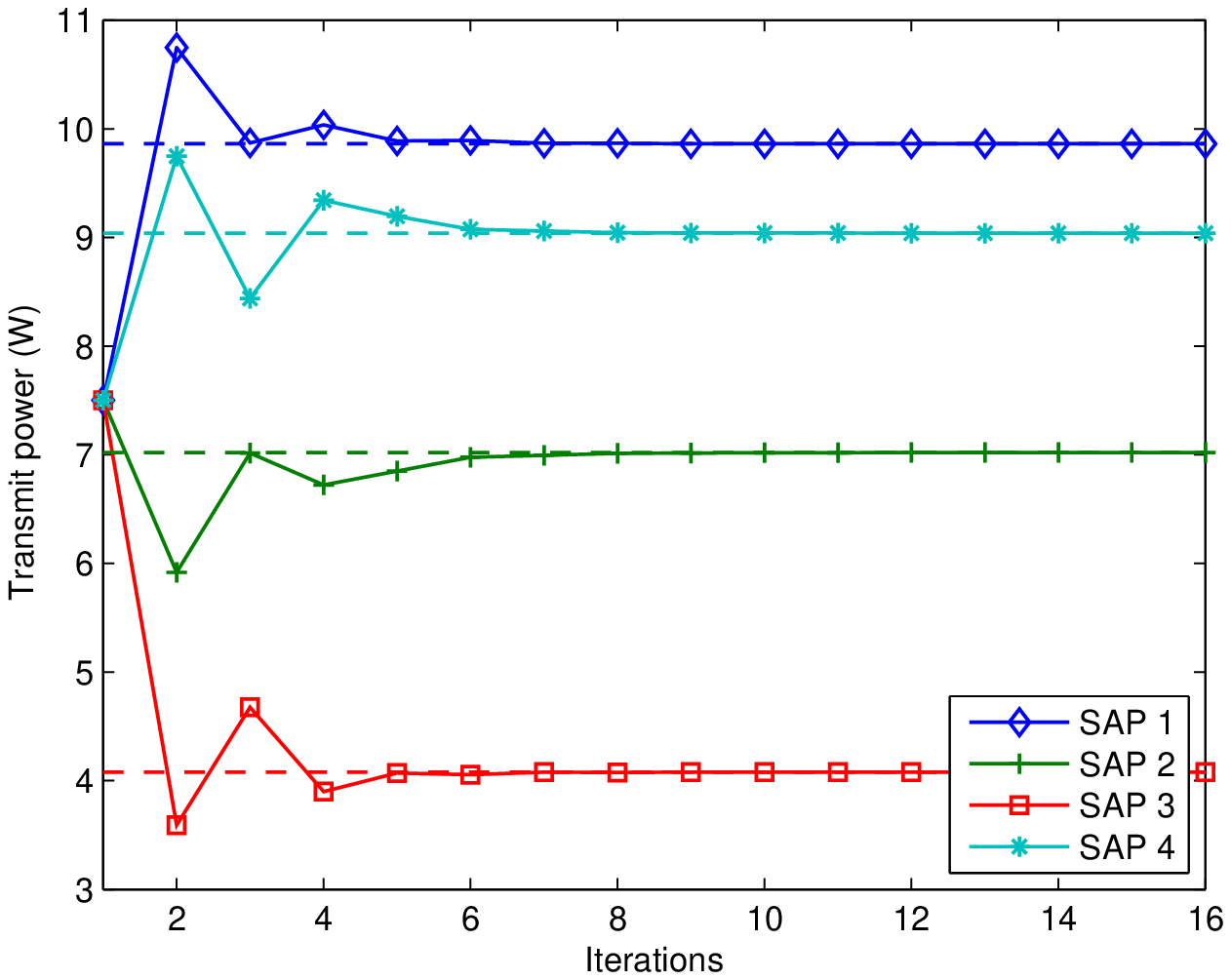}
    \label{fig:sim_FiniteSys_Converg_p}}
    \subfigure[Convergence result of dual variable $\nu_i$, $\forall i$]
    {\includegraphics[width=0.48\textwidth]{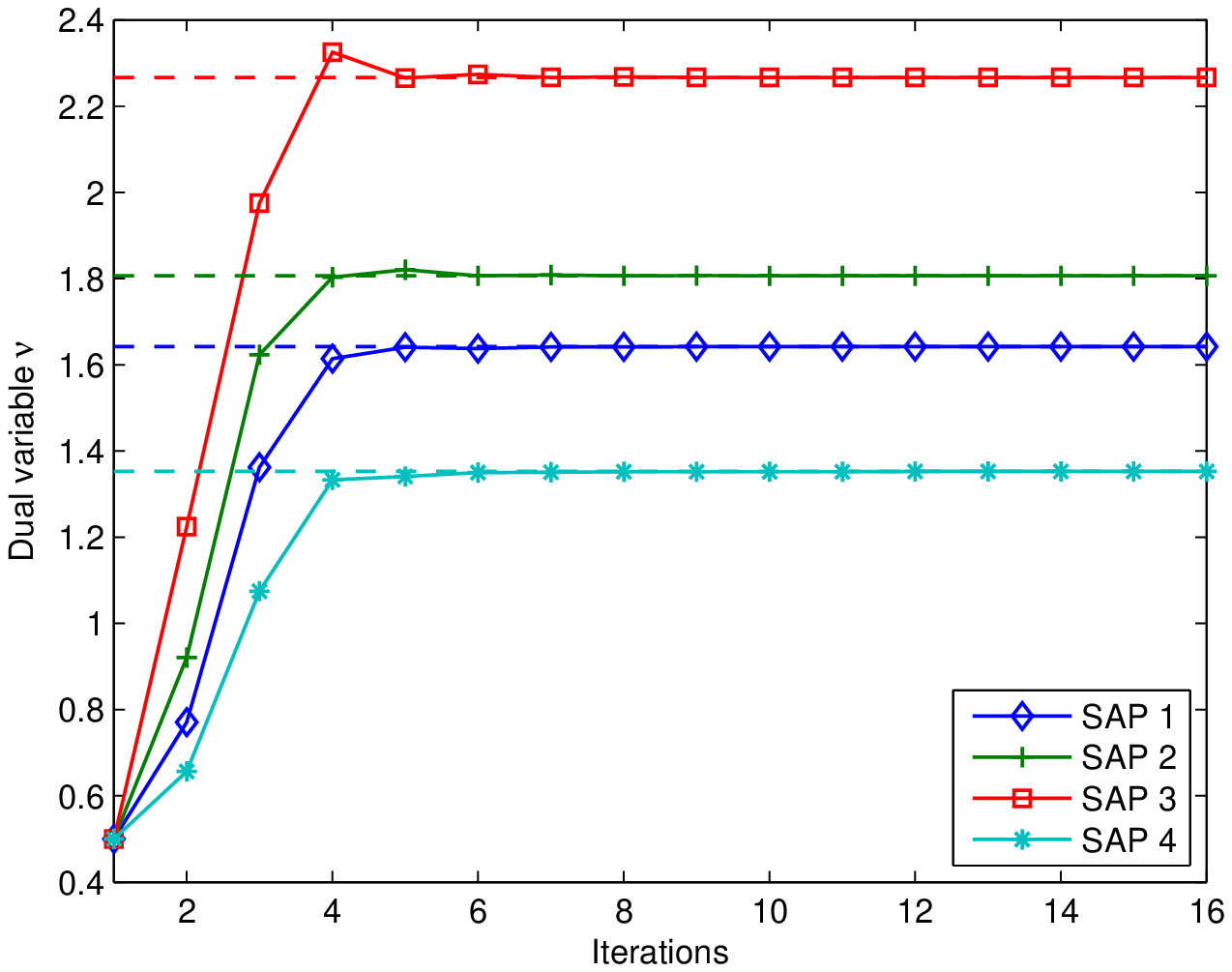}
    \label{fig:sim_FiniteSys_Converg_nu}}%
    \caption{Convergence results for finite system iterative algorithms. The solid lines show the convergence of Algorithm~\ref{alg:Alg1}, and the dashed lines show the final results obtained by AUC. 
}\label{fig:sim_FiniteSys_Converg}
\end{figure*}
The convergence behavior of Algorithm~\ref{alg:Alg1} is shown in Fig.~\ref{fig:sim_FiniteSys_Converg}. Algorithm~\ref{alg:Alg1} converges to the same result obtained by AUC in less than 8 iterations. Fig.~\ref{fig:sim_FiniteSys_Converg_p} shows the convergence results of $\CB{p_i}$, and we initialize all $p_i$ with the same value. Fig.~\ref{fig:sim_FiniteSys_Converg_nu} shows the convergence results of $\CB{\nu_i}$. After 8 iterations, the difference between Algorithm~\ref{alg:Alg1} and the final result of AUC is negligible.




\begin{figure}
  \centering
  \includegraphics[width=0.48\textwidth]{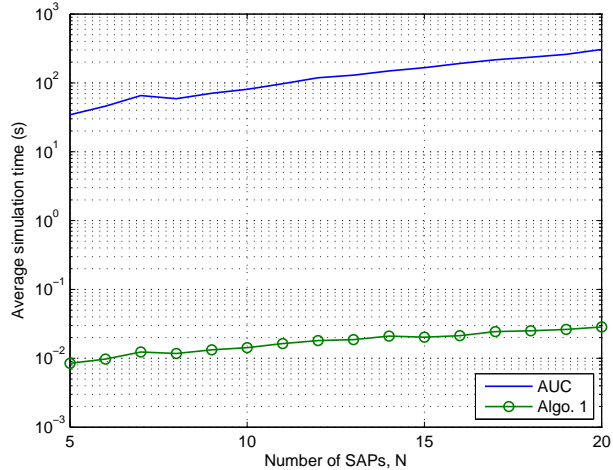}\\
  \caption{Comparison of average simulation time for each channel realization using Algorithm~\ref{alg:Alg1} and AUC. 
}\label{fig:sim_FiniteSys_Complexity_M3}
\end{figure}

In order to compare the computational complexity of Algorithm~\ref{alg:Alg1} and AUC, we perform simulations on a PC with a Core 2 Duo CPU of internal clock 3.00GHz and 4GB RAM. Twenty channel realizations are performed for each algorithm. The number of SAPs varies from $N = 5$ to 20. The SINR requirements at the SAPs are set to $\gamma_i = 7.78$dB, $\forall i$. The stopping criteria for both algorithms are such that the maximum uplink power difference between neighboring iterations should be smaller than $\epsilon = 10^{-5}$.
The average simulation time for each channel realization using Algorithm~\ref{alg:Alg1} and AUC is shown in Fig.~\ref{fig:sim_FiniteSys_Complexity_M3}. We observe that the average simulation time using AUC is 5000 to 7000 times longer than that of Algorithm~\ref{alg:Alg1}. \addedv{This is because solving the GP problem~\eqref{opt:GP} in each iteration requires significant amount of time using standard interior point algorithms compared to the iterative Algorithm~\ref{alg:Alg1}.} Moreover, the ratio of simulation time increases with the number of SAPs $N$. This shows that the proposed iterative algorithm reduces huge amount of computation to solve~\eqref{opt:DL_L1} without invoking standard convex optimization software.

\subsection{Convergence Behavior of Large System Iterative Algorithm}

\begin{figure*}
\centering
    \subfigure[Transmit powers $p_i/M$, $\forall i$]
    {\includegraphics[width=0.48\textwidth]{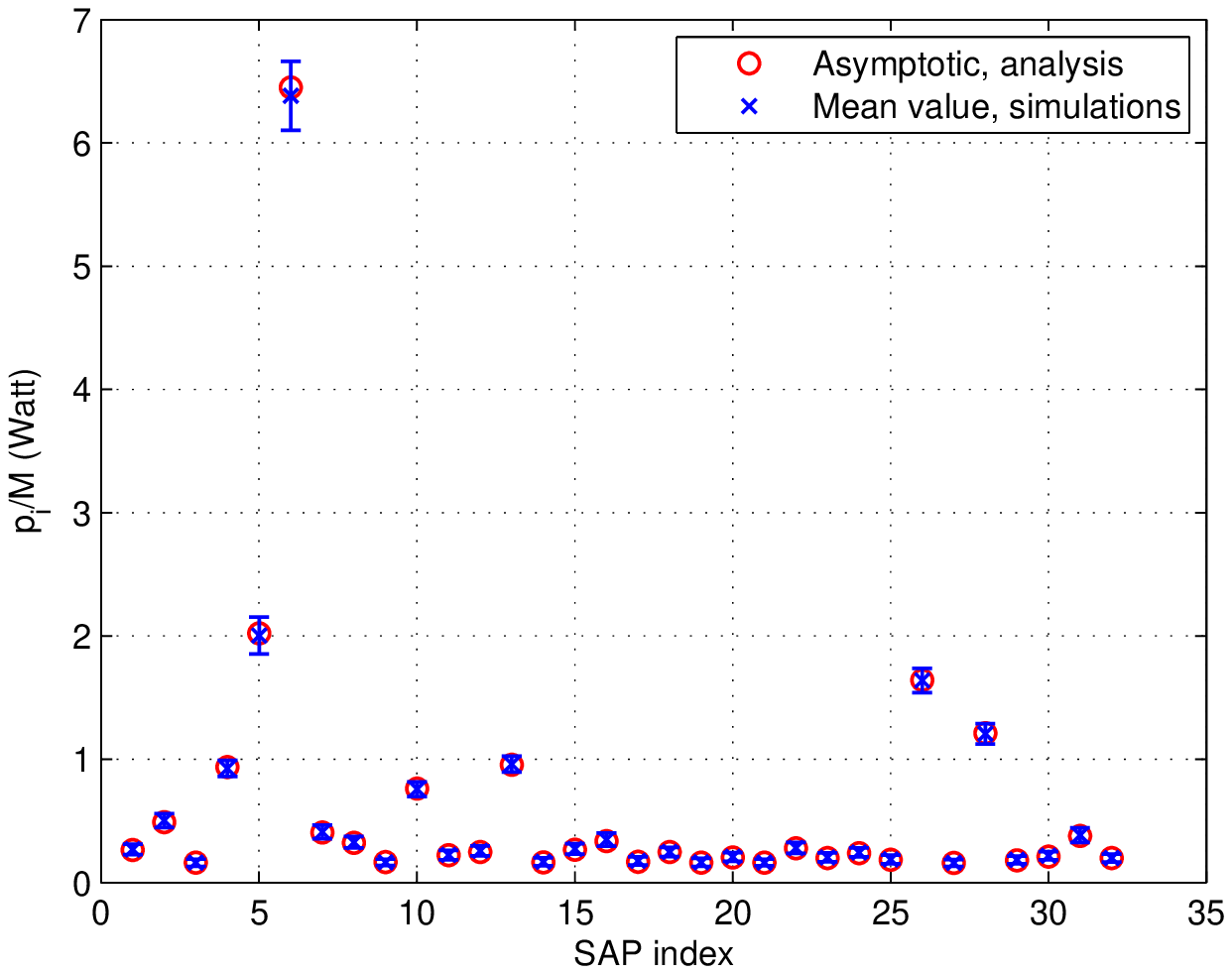}
    \label{fig:sim_RMT_Converg_p}}
    \subfigure[Values of $\nu_i$, $\forall i$]
    {\includegraphics[width=0.485\textwidth]{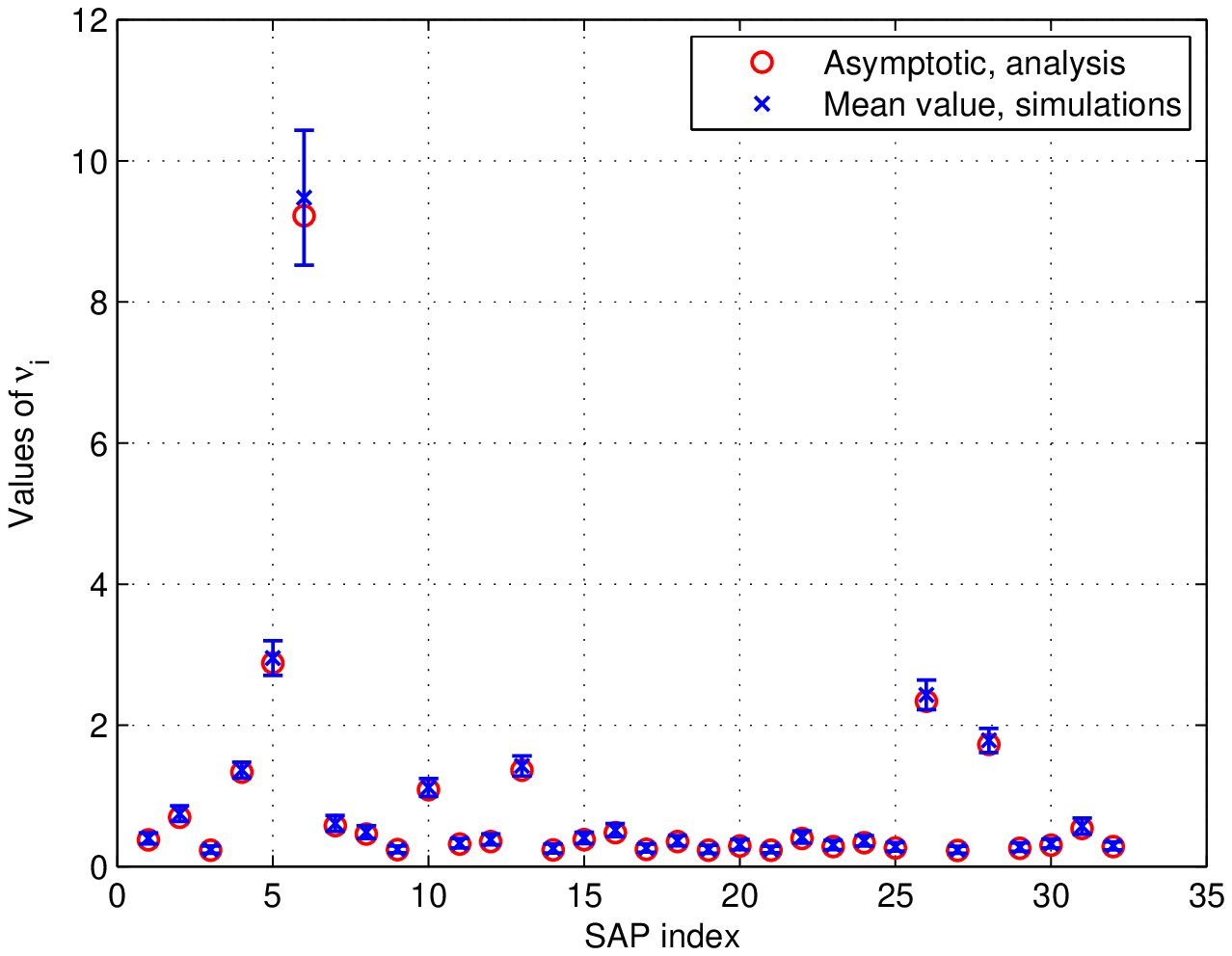}
    \label{fig:sim_RMT_Converg_nu_lin}}%
    \caption{Comparison of final results in $\vec{p}$ and $\bm{\nu}$ using large system analysis and Monte Carlo simulations. Monte Carlo simulations are performed in 100 i.i.d. channel realizations, and their results are represented using error bars. 
}\label{fig:sim_RMT_Converg}
\end{figure*}


Numerical simulations are carried out to verify the iterative algorithm for large systems. In the simulation setup, the WBH has $M = 64$ antennas and there are $N = 32$ single antenna SAPs. The weights of power are set to $w_i = 1$, $\forall i$. The power constraint at the WBH is $P = 20$ Watt and the noise variance at each SAP is normalized to $n_i = 1$ Watt, $\forall i$. The SINR requirement at each SAP is set to $\gamma_i = 6.02$dB, $\forall i$. The large-scale fading coefficients $\CB{d_i}$ between the WBH and the SAPs are assigned some randomly generated positive values, and they are fixed in the simulation.

\changed{In Fig.~\ref{fig:sim_RMT_Converg}, we compare the results obtained by large system analysis with those obtained by Monte Carlo simulations. For the large system analysis, we calculate the transmit power $\CB{p_i/M}$ and dual values of $\CB{\nu_i}$ according to Algorithm~\ref{alg:Alg2}. Those values are the theoretical deterministic quantities that $\CB{p_i/M}$ and $\CB{\nu_i}$ converge to when the system dimensions go to infinity, and they are shown in vertical bars for each SAP in Fig.~\ref{fig:sim_RMT_Converg_p} and Fig.~\ref{fig:sim_RMT_Converg_nu_lin}, respectively. For the Monte Carlo simulations, 100 i.i.d. channel realizations are carried out. For each channel realization, the transmit power $\CB{p_i/M}$ and dual values of $\CB{\nu_i}$ are obtained using Algorithm~\ref{alg:Alg1}. For each SAP $i$, the values of $p_i/M$ and $\nu_i$ obtained by Monte Carlo simulations are shown in error bars, where the mean value is indicated by a cross sign and the distance above and below the mean value denotes the standard deviation. We observe that the mean values accurately match the results obtained by large system analysis and the standard deviations are usually very small in both Fig.~\ref{fig:sim_RMT_Converg_p} and Fig.~\ref{fig:sim_RMT_Converg_nu_lin}. This observation shows that $\CB{p_i/M}$ and $\CB{\nu_i}$ really converge to the deterministic quantities predicted by large system analysis irrespective of the actual channel realization as the system dimensions become large. Instead of the instantaneous channel information, the large system analysis Algorithm~\ref{alg:Alg2} utilizes the large-scale fading coefficients $\CB{d_i}$ as the channel inputs and requires infrequent updates.
}

\subsection{Cellular Network Simulations}
\addedv{We perform numerical simulations for SAP admission control using a macrocell network setup with multiple small cells.} The WBH is located at the center of the cell. We assume the MBS is co-located with the WBH and has the same number of antennas. The cell radius is 1km. The transmit antenna gain at the WBH is 5dB. The pathloss model from the WBH to the SAPs is
\begin{equation}
    L (\text{dB}) = 128+37.6 \cdot \log_{10}D,
\end{equation}
where $D$ represents the distance between the WBH and the SAP in the unit of km. The log-normal shadowing parameter is 10dB. The bandwidth of the wireless backhaul is 10MHz. The WBH transmit power constraint is 30dBm. The noise variance at each SAP is -93.98dBm. The cell radius of each small cell is 30m.
We assume the SAPs are randomly and uniformly distributed within the cell.
\addedv{The channel pathloss, shadowing parameters, transmit power constraints and antenna gains are based on~\cite{3GPPTS36814}.}
Table~\ref{tab:Parameters} shows a summary of the above simulation parameters.

\begin{table}
  \renewcommand{\arraystretch}{1.3}
  \centering
  \caption{Summary of simulation parameters}\label{tab:Parameters}
  \begin{tabular}{|c|c|}
    \hline
    Number of macrocells & single \\ \hline
    Cell radius & 1km \\ \hline
    Small cell radius & 30m \\ \hline
    WBH transmit antenna gain & 5dB \\ \hline
    WBH transmit power constraint & 30dBm \\ \hline
    Log-normal shadowing & 10dB \\ \hline
    Transmission spectrum for backhaul & 10MHz \\ \hline
    Noise variance & -93.98dBm \\ \hline
    SAP location distribution & uniform \\ \hline
    Number of antennas per SAP & 1 \\ \hline
    Number of antennas per user & 1 \\
    \hline
  \end{tabular}
\end{table}

\begin{figure*}
\centering
    \subfigure[Average number of admitted SAPs]
    {\includegraphics[width=0.48\textwidth]{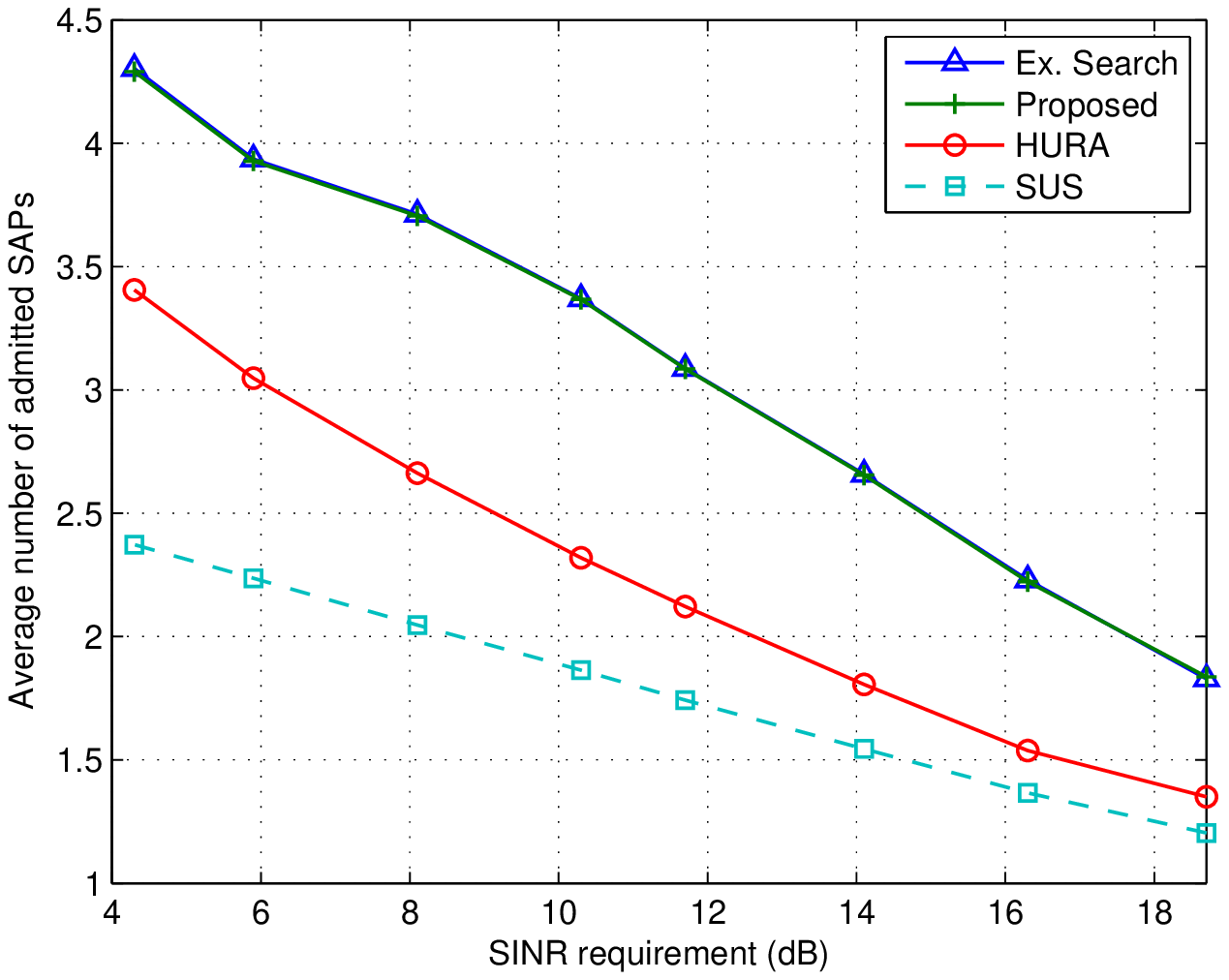}
    \label{fig:comp_Algo_SAP_n}}
    \subfigure[Transmit power $\sum_i p_i$]
    {\includegraphics[width=0.485\textwidth]{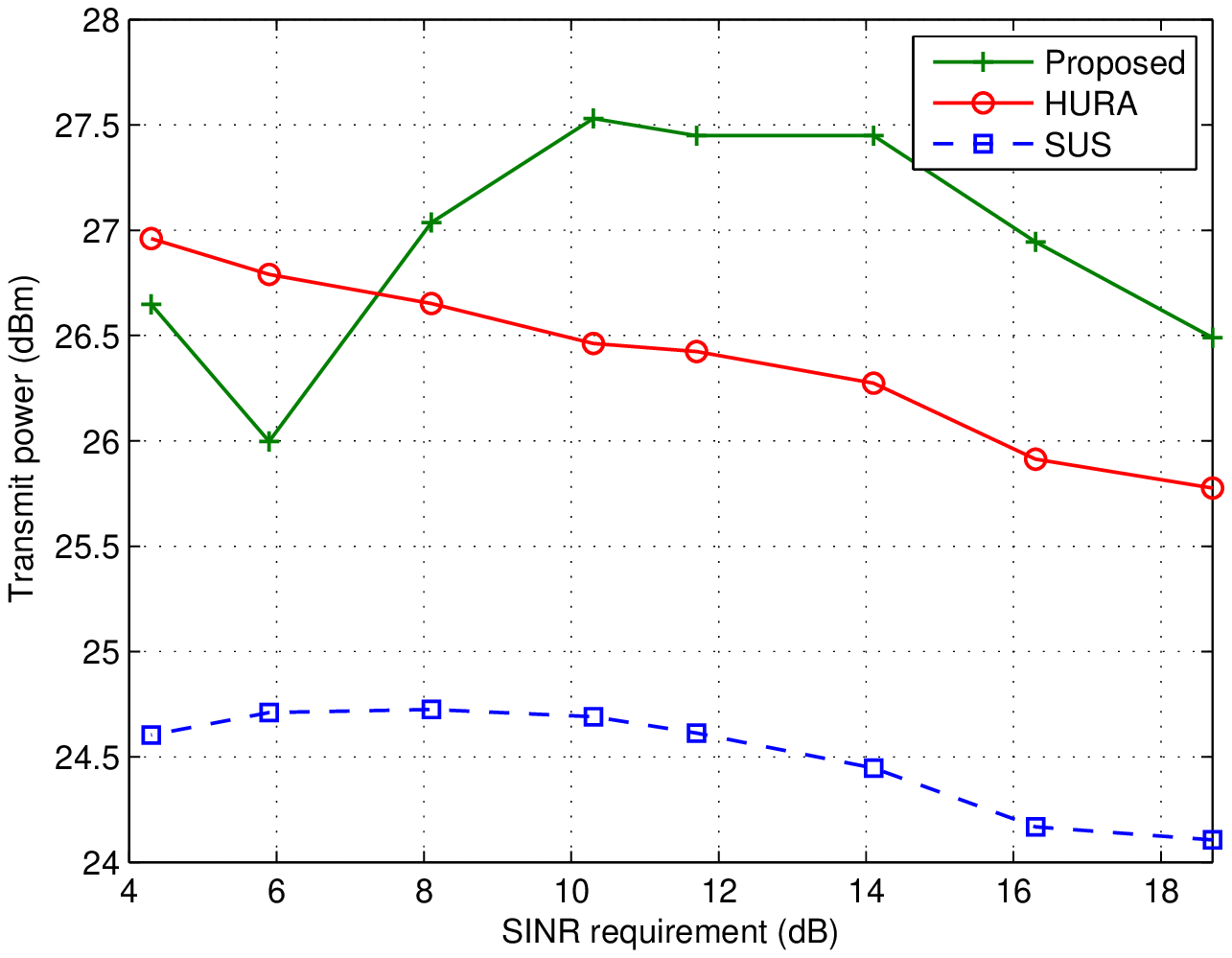}
    \label{fig:comp_Algo_SAP_p}}%
    \caption{Cellular network simulations for finite system SAP admission. 
}\label{fig:comp_Algo_SAP}
\end{figure*}

In the numerical simulations for finite system SAP admission, we compare the proposed SAP admission control method, which employs Algorithm~\ref{alg:Alg1} and iterative SAP removal, to other two commonly used methods in literature: the Lagrange duality based \emph{heuristic removal user admission} (HRUA) algorithm~\cite{stridh2006system} and the \emph{semiorthogonal user selection} (SUS) algorithm~\cite{yoo2006optimality}. We also compare the results with the exhaustive search (ES) method. The ES method obtains the maximum number of admitted SAPs by searching through all possible choices of SAPs and chooses the set of SAPs with the maximum cardinality. Even though the ES method produces the optimum results for~\eqref{opt:DL_L0}, the computational load of the ES method is very high and it is only used as a benchmark for comparing different methods. We consider a network where the WBH has $M = 4$ antennas and there are $N = 12$ SAPs in the network. We generate 60 channel realizations for each SAP location layout and 20 different SAP location layouts are performed in simulations.
The results for the considered methods are shown in Fig.~\ref{fig:comp_Algo_SAP}. The average number of admitted SAPs for each channel realization is shown in Fig.~\ref{fig:comp_Algo_SAP_n}. We observe that the average number of admitted SAPs by our proposed iterative method is nearly identical to that obtained by the optimum ES method. The HURA algorithm supports one less SAP compared to the ES method on average. The results achieved by the SUS algorithm is the worst, which is 2 to 1.2 less than the ES method. The transmit power $\sum_i p_i$ is shown in Fig.~\ref{fig:comp_Algo_SAP_p}. Our proposed algorithm generally has higher power compared to HURA and SUS, but all the considered algorithms satisfy the transmit power constraint $P$.



\begin{table}
  \renewcommand{\arraystretch}{1.3}
  \centering
  \caption{Comparison of SAP admission control schemes with unequal SINR requirements}\label{tab:compSAP}
  \begin{tabular}{|c|c|c|c|c|}
    \hline
     & Ex. Search & Proposed & HURA & SUS \\ \hline
    Average admitted SAPs / users & 3.49 / 3.42 & 3.29 / 3.21 & 1.78 / 1.75 & 1.80 / 1.78 \\ \hline
    Average WBH transmit power (dBm) & 28.73 & 27.34 & 25.66 & 24.61 \\ \hline
    Average SAP / user sum rate (b/s/Hz) & 10.97 / 10.69 & 11.16 / 10.79 & 6.92 / 6.80 & 6.46 / 6.35 \\
    \hline
  \end{tabular}
\end{table}
\changedv{
We also perform simulations using the cellular network parameters in Table~\ref{tab:Parameters} for the case when the QoS requirements at different SAPs are unequal and change with time. In the simulation, each SAP is responsible for serving one user. We generate 60 different SAP location layouts. For each SAP location layout, 20 different user locations in each SAP are considered.
In each time slot, the SINR requirement of each user is drawn randomly from 4.3dB to 18.7dB. The simulation results for different SAP admission schemes are shown in Table~\ref{tab:compSAP}. After the admission of SAPs, we use the max-min SINR algorithm~\cite{cai2012maxmin} to verify whether the users in the admitted small cells can be simultaneously supported for their SINR requirements. If not, we apply the user admission control discussed in Section~\ref{subsec: UserAdmission} to select the maximum set of users to satisfy their QoS requirements within the admitted small cells. The user rates are obtained from the finally admitted users.
For the ES method, there may be more than one set of SAPs that have the maximum cardinality. In that case, we randomly choose one set of SAPs from them due to complexity issues. We observe from Table~\ref{tab:compSAP} that the average number of admitted SAPs for the proposed scheme is very close to the optimum results obtained by the ES method, which far outperforms the HURA and SUS methods. The average user sum rate achieved by the proposed scheme is also 58.7\% and 70\% higher than the HURA and SUS methods, respectively. In Table~\ref{tab:Parameters}, the number of finally admitted users is quite close to that of the admitted SAPs. This shows that the small cell user admission control only needs to be performed rarely.}


\changedv{The simulation results comparing the user throughput with and without SAPs in a cellular network is shown in Table~\ref{tab:compMS}. We use the simulation parameters in Table~\ref{tab:Parameters} and the simulation setup is similar to that described for Table~\ref{tab:compSAP}. We compare two cases: one is that the WBH performs SAP admission and transmit data to the admitted SAPs using wireless backhaul. Then the admitted SAPs transmit those data to their users using in-band channel taking into account the intercell interference. If the admitted users cannot be simultaneously supported for the given SINR requirements, user admission control discussed in Section~\ref{subsec: UserAdmission} is applied.
The other is that the users are served directly by the MBS using in-band channel. In both cases, the transmit power of in-band channel is the same. In Table~\ref{tab:compMS}, we observe that the user throughput with SAP admission is about 8.3 times that without SAPs. This is because the deployment of SAPs significantly reduces the distance between the transmitters and users. Without using small cells, the signal strength at the users will be much weaker when transmitted from the MBS. This shows that small cells can boost the system throughput considerably.} 
\begin{table}
  \renewcommand{\arraystretch}{1.3}
  \centering
  \caption{Comparison of user throughput with and without SAPs}\label{tab:compMS}
  \begin{tabular}{|c|c|c|c|c|}
    \hline
     & with SAPs & without SAPs \\ \hline
    Average admitted (SAPs) / users & 3.31 / 3.21 & 0.43 \\ \hline
    Average (SAP) / user sum rate (b/s/Hz) & 11.17 / 10.75 & 1.30 \\
    \hline
  \end{tabular}
\end{table}


\begin{figure*}
\centering
    \subfigure[Average number of admitted SAPs]
    {\includegraphics[width=0.48\textwidth]{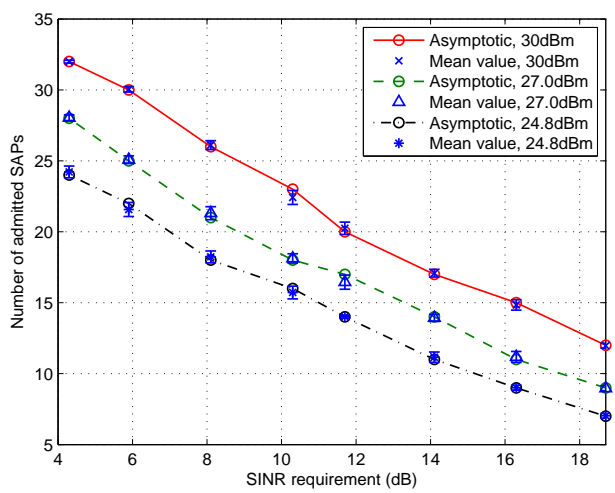} 
    \label{fig:sim_RMT_Cellular}}
    \subfigure[Min. SINR for SAPs admitted using large system analysis]
    {\includegraphics[width=0.492\textwidth]{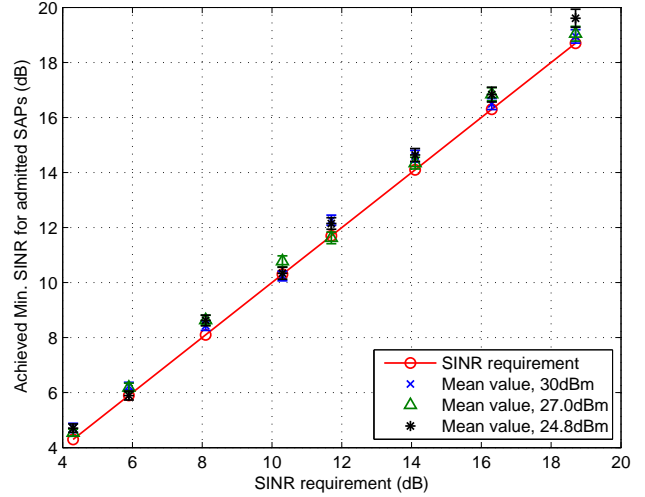} 
    \label{fig:sim_RMT_Cellular_SINR}}%
    \caption{Comparing results using large system analysis and Monte Carlo simulations. Monte Carlo simulations are performed in 100 i.i.d. channel realizations, and their results are represented using error bars. 
}\label{fig:sim_RMT}
\end{figure*}
\changed{The simulations for large systems are carried out employing cellular network parameters of Table~\ref{tab:Parameters}. In the simulations, the WBH is equipped with $M = 64$ antennas and there are $N = 32$ single antenna SAPs. The weights of power are set to $w_i = 1$, $\forall i$. The power constraints are set to $P = 30$dBm, 27.0dBm, and 24.8dBm. We compare the simulation results using large system analysis and those using Monte Carlo simulations in Fig.~\ref{fig:sim_RMT}.
Fig.~\ref{fig:sim_RMT_Cellular} compares the number of admitted SAPs under given power and SINR constraints. The asymptotic results are shown in circles, and they are obtained by the SAP admission control method employing Algorithm~\ref{alg:Alg2} and iterative SAP removal using large-scale fading coefficients. Those asymptotic results are compared to the Monte Carlo simulation results, where 100 i.i.d. channel realizations are carried out. For each channel realization, the Monte Carlo simulation performs SAP admission employing Algorithm~\ref{alg:Alg1} and iterative SAP removal. Those Monte Carlo results are shown in error bars. We observe from Fig.~\ref{fig:sim_RMT_Cellular} that the standard deviations of the Monte Carlo results are very small, which shows the Monte Carlo results tend to be deterministic values. Such deterministic values are accurately predicted by the asymptotic results using large system analysis.
In Fig.~\ref{fig:sim_RMT_Cellular_SINR}, we show the achieved minimum SINR by the SAPs that are selected employing the large system analysis. For the given SINR constraint and large-scale fading coefficients, the set of admitted SAPs is selected using the SAP admission control method employing Algorithm~\ref{alg:Alg2} and iterative SAP removal. The max-min SINR algorithm~\cite{cai2011unified} is utilized for each channel realization to maximize the minimum SINR for that set of admitted SAPs over 100 i.i.d. channel realizations. The error bars show the achieved SINR values for different channel realizations, which are very close or above the SINR requirements. Fig.~\ref{fig:sim_RMT_Cellular_SINR} shows that the admitted SAPs can satisfy the SINR requirements in nearly all channel realizations even though the SAPs are selected using large system analysis.
}

\section{Conclusion}\label{sec:Conclusions}
We considered the problem of SAP admission control in a heterogeneous cellular network using wireless backhaul. In order to divert the users from the macrocell to small cells, \added{as well as to minimize the total cost of building backhaul to serve small cells in the network}, the WBH needs to simultaneously serve as many SAPs as possible under given power and SINR constraints. Such a problem is combinatorial and NP-hard. We applied $\ell_1$-norm relaxation and proposed an iterative algorithm to solve the relaxed problem. The local convergence property of the iterative algorithm was proved. Based on the solution of the $\ell_1$-relaxed problem, the SAPs were iteratively removed until all the remaining SAPs can satisfy the power and SINR constraints.
We also proposed a large system iterative algorithm using random matrix theory. Such a algorithm only requires large-scale channel coefficients to perform SAP admission control for large systems irrespective of instantaneous channel information.
Simulations showed that the finite system iterative algorithm achieved near optimum results and the large system iterative algorithm predicted the Monte Carlo simulation results accurately.


\appendices

\section{Proof of Lemma~\ref{lem:UL_DL}}\label{apx:Proof_UL_DL}
For any set of given beamformers $\CB{\vec{u}_i}$, according to \cite[Proposition 27.2]{Boche2005dualityBook}, both the uplink and downlink have the same SINR feasible region under the power constraint $P$ with the uplink and downlink SINR definition of \eqref{eq:SINR_U} and \eqref{eq:SINR_D}, respectively. The target SINRs $\gamma_i/(1+x_i)$, $\forall i$, are feasible in the downlink if and only if the same targets are feasible in the uplink. Therefore, the same set of $\vec{x}$ is feasible for \eqref{opt:DL_L1} and \eqref{opt:UL_L1} under the power constraint $P$. Since this is true for any set of beamformers $\CB{\vec{u}_i}$, the optimum solution of $\CB{\vec{u}_i}$ and $\vec{x}$ in \eqref{opt:DL_L1} and \eqref{opt:UL_L1} are the same. For given $\CB{\vec{u}_i}$ and $\vec{x}$, the mapping between uplink power $\vec{q}$ and downlink power $\vec{p}$ that achieves the same SINRs can be obtained by the uplink-downlink power mapping~\cite{bengtsson2001optimal}.

\section{Proof of Lemma~\ref{lem:MMSE}}\label{apx:Proof_MMSE}
Consider the uplink power allocation $\vec{q}^{\star}$, it is known~\cite{van2002optimum} that the uplink SINR \eqref{eq:SINR_U} is maximized by the MMSE receiver, i.e., $\SINR_i^{\rm{U}} \RB{\vec{q}^{\star}, \CB{\vec{u}_i^{\rm{MMSE}}}} \ge \SINR_i^{\rm{U}} \RB{\vec{q}^{\star}, \CB{\vec{u}_i}}$. If $\vec{u}_i \ne \vec{u}_i^{\rm{MMSE}}$ in the solution of \eqref{opt:UL_L1}, substituting $\vec{u}_i$ with $\vec{u}_i^{\rm{MMSE}}$ will also satisfy \eqref{opt:UL_L1}. Therefore, $\CB{\vec{u}_i^{\rm{MMSE}}}$ must be the solution of optimum receive beamformers in \eqref{opt:UL_L1}. The corresponding SINR can be obtained by substituting~\eqref{eq:mmse_BF} into~\eqref{eq:SINR_U}.

\section{Proof of Lemma~\ref{thm1}}\label{apx:Proof_KKT}
The Lagrangian of \eqref{opt:GP} can be expressed as
\begin{equation}
    \mathcal{L}(\vec{\qt}, \CB{\vec{u}_i}, \vec{x}) = \sum_i (1-\alpha_i) x_i + \sum_i \nu_i \RB{\log\frac{ \gamma_i \RB{\sum_{j \ne i}G_{ji}e^{\qt_j} + w_i} }{G_{ii}e^{\qt_i}} - \log(1+x_i)} + \mu \RB{\sum_i n_i e^{\qt_i} - P}.
\end{equation}

According to the KKT conditions, we have
\begin{IEEEeqnarray}{rCl}
  \frac{\partial \mathcal{L}}{\partial x_i} &=& 1 - \alpha_i - \frac{\nu_i}{1+x_i} = 0, \quad \forall i = 1, \cdots, N; \\
  \frac{\partial \mathcal{L}}{\partial \qt_i} &=& -\nu_i + \sum_{j \ne i} \nu_j \frac{G_{ij}e^{\qt_i}}{\sum_{k \ne j}G_{kj}e^{\qt_k} + w_j} + \mu n_i e^{\qt_i} = 0, \quad \forall i.
\end{IEEEeqnarray}
Therefore, we have
\begin{IEEEeqnarray}{rCl}
  x_i  &=& \max\RB{\nu_i - 1, 0}, \quad \forall i; \label{eq:KKT1} \\
  \nu_i &=& \RB{\sum_{j \ne i} \frac{\nu_j G_{ij}}{\sum_{k \ne j}G_{kj}e^{\qt_k} + w_j} + \mu n_i} e^{\qt_i}, \quad \forall i. \label{eq:KKT2}
\end{IEEEeqnarray}

Because $e^{\qt_i} > 0$, any $\nu_i = 0$ requires $\mu = 0$ and $\nu_j = 0$, $\forall j = 1, \cdots, N$, simultaneously. By abandoning this trivial solution, we have $\mu > 0$ and $\nu_i > 0$, $\forall i$. According to the complementary slackness conditions~\cite{Boyd04ConvexOptimization} and substituting $e^{\qt_i} = q_i/M$, we have
\begin{IEEEeqnarray}{rCl}
  \frac{(1+x_i) G_{ii}q_i}{M \gamma_i} &=& \sum_{j \ne i}\frac{G_{ji}q_j}{M} + w_i, \quad \forall i; \label{eq:KKT3} \\
  \sum_i \frac{n_i q_i}{M} &=& P. \label{eq:KKT4}
\end{IEEEeqnarray}

Substituting \eqref{eq:KKT3} into \eqref{eq:KKT2}, we have
\begin{equation}\label{eq:KKT5}
    \frac{M \nu_i}{q_i} = \sum_{j \ne i}\frac{M G_{ij} \gamma_j \nu_j}{(1+x_j) G_{jj} q_j} + \mu n_i, \quad \forall i.
\end{equation}
Therefore
\begin{equation}\label{eq:KKT6}
    \frac{(1+x_i)}{\gamma_i}G_{ii}\frac{M \gamma_i\nu_i}{(1+x_i) G_{ii} q_i \mu} = \sum_{j \ne i}G_{ij}\frac{M \gamma_j \nu_j}{(1+x_j) G_{jj} q_j \mu} + n_i, \quad \forall i.
\end{equation}
We define
\begin{equation}\label{eq:pj}
  \frac{p_i}{M} \triangleq \frac{M \gamma_i\nu_i}{(1+x_i) G_{ii} q_i \mu}, \quad \forall i
\end{equation}
and we have
\begin{equation}\label{eq:KKT7}
    \frac{(1+x_i) G_{ii}p_i}{M \gamma_i} = \sum_{j \ne i}\frac{G_{ij}p_j}{M} + n_i, \quad \forall i.
\end{equation}
Multiply both sides of \eqref{eq:KKT3} with $p_i$ and sum them up for all $i$. We also multiply both sides of \eqref{eq:KKT7} with $q_i$ and sum them up for all $i$. Because $\sum_i \sum_{j \ne i} G_{ji} q_j p_i = \sum_i \sum_{j \ne i} G_{ij} q_i p_j$, we have 
\begin{equation}\label{eq:KKT9}
    \sum_i \frac{w_i p_i}{M} = \sum_i \frac{n_i q_i}{M} = P.
\end{equation}
Eq.~\eqref{eq:KKT7} shows the power allocation $\vec{p}$ defined in \eqref{eq:pj} achieves the same SINR as the uplink $\vec{q}$. Eq.~\eqref{eq:KKT9} shows $\vec{p}$ satisfies the same power constraint as $\vec{q}$. When the variables on the right-hand-side of \eqref{eq:pj} are the optimum primal and dual solutions of \eqref{opt:UL_L1}, the power $\vec{p}$ defined in \eqref{eq:pj} corresponds to the optimum downlink power that solves \eqref{opt:DL_L1} according to Lemma~\ref{lem:UL_DL}.

%

\section{Proof of Theorem~\ref{thm2}}\label{apx:Proof_Converge}
We only show the proof assuming that $\nu_i^{\star} \ge 1$, $\forall i$, at the optimum solution of \eqref{opt:UL_L1}. The proof that some $\nu_i^{\star} < 1$ can be obtained likewise.
Since we consider points that are sufficiently close to the optimum solution, we assume the beamformers $\CB{\vec{u}_i}$ to be sufficiently close to the optimum beamformers $\CB{\vec{u}_i^{\star}}$. Then the equivalent channel is $G_{ij} = \abs{\vec{h}_i^H \vec{u}_j^{\star}}^2$, $\forall i,j$, and it can be assumed to be fixed. The updates in Algorithm~\ref{alg:Alg1} boil down to the updates of $q_i$ and $\nu_i$, $\forall i$. We define a vector $\bm{\omega} \in \Real_{+}^N$ and a matrix $\mat{F} \in \Real_{+}^{N \times N}$, where $\omega_i = M w_i \gamma_i / G_{ii}$, $\forall i = 1, \cdots, N$, and
\begin{equation}
  F_{ij} =
  \left\{
    \begin{array}{ll}
      \frac{G_{ij} \gamma_j}{G_{jj}}, & \hbox{if $i \ne j$;} \\
      0, & \hbox{if $i = j$.}
    \end{array}
  \right.
\end{equation}
Furthermore, we introduce a vector $\vec{y} = \SB{y_1, \cdots, y_N}^T$, where $y_i = \nu_i/q_i$, $\forall i$. Therefore, $\bm{\nu} = \vec{y} \circ \vec{q}$.

The updates of $\vec{q}$ in \eqref{step:q} is actually obtained from the KKT condition \eqref{eq:KKT3_} using MMSE receivers $\CB{\vec{u}_i}$. Since we assume $\bm{\nu}^{\star} \ge \bm{1}$ and $\vec{u}_i \approx \vec{u}_i^{\star}$ here, the updates of $\vec{q}$ can be expressed as
\begin{IEEEeqnarray}{rCl}
  \vec{q}^{(m+1)} &=& \diag\RB{\bm{\nu}^{(m)}}^{-1} \RB{\mat{F}^T \vec{q}^{(m)} + \bm{\omega}} \\
  &=& \diag\RB{\vec{q}^{(m)}\circ\vec{y}^{(m)}}^{-1} \RB{\mat{F}^T \vec{q}^{(m)} + \bm{\omega}}.
\end{IEEEeqnarray}
By substituting \eqref{step:lambda} and \eqref{step:p} into \eqref{step:nu}, the updates of $\vec{y}$, which is obtained by $\vec{y} = \bm{\nu} \circ \vec{q}^{-1}$, can be expressed as
\begin{IEEEeqnarray}{rCl}
  \vec{y}^{(m+1)} &=& \mat{F}\cdot\RB{\vec{q}^{(m)}}^{-1} + \frac{1}{MP}\vec{n} \cdot \RB{\bm{\omega}^T\RB{\vec{q}^{(m)}}^{-1}}\\
  &=& \RB{\mat{F} + \frac{1}{MP}\vec{n}\bm{\omega}^T} \RB{\vec{q}^{(m)}}^{-1}.
\end{IEEEeqnarray}



By dropping the time indices and letting $\vec{z} = \SB{\vec{q}^T, \vec{y}^T}^T$, the fixed-point updates of $\vec{z}$ can be expressed as
\begin{equation}\label{eq:Mapping}
  T(\vec{z}) = \left(
    \begin{array}{c}
      f_1(\vec{q}, \vec{y}) \\
      f_2(\vec{q}, \vec{y}) \\
    \end{array}
  \right) =
  \left(
    \begin{array}{c}
      \diag(\vec{q}\circ\vec{y})^{-1} (\mat{F}^T \vec{q} + \bm{\omega}) \\
      \RB{\mat{F} + \frac{1}{MP}\vec{n}\bm{\omega}^T} \vec{q}^{-1} \\
    \end{array}
  \right).
\end{equation}
Its Jacobian matrix can be written as
\begin{IEEEeqnarray}{rCl}
  \mat{J} &=& \left(
                \begin{array}{cc}
                  \partial f_1/\partial \vec{q}^T & \partial f_1/\partial \vec{y}^T \\
                  \partial f_2/\partial \vec{q}^T & \partial f_2/\partial \vec{y}^T \\
                \end{array}
              \right)
   \\
   &=& \left(
         \begin{array}{cc}
           \diag(\vec{q}\circ\vec{y})^{-1}\mat{F}^T & \mat{0} \\
           \mat{0} & \mat{0} \\
         \end{array}
       \right)
   - \mat{E} \cdot \left(
                     \begin{array}{cc}
                       \diag(\vec{q}\circ\vec{y})^{-2} & \mat{0} \\
                       \mat{0} & \diag(\vec{q}\circ\vec{y})^{-2} \\
                     \end{array}
                   \right)
\end{IEEEeqnarray}
where
\begin{equation}\label{eq:E}
  \mat{E} = \left(
              \begin{array}{cc}
                \diag(\mat{F}^T \vec{q} + \bm{\omega})\diag(\vec{y}) & \diag(\mat{F}^T \vec{q} + \bm{\omega})\diag(\vec{q}) \\
                \RB{\mat{F} + \frac{1}{MP}\vec{n}\bm{\omega}^T}\diag(\vec{y})^2 & \mat{0} \\
              \end{array}
            \right).
\end{equation}

At the optimum solution, we have
\begin{IEEEeqnarray}{rCl}
  \vec{q}^{\star} &=& \diag(\vec{q}^{\star} \circ \vec{y}^{\star})^{-1} (\mat{F}^T \vec{q} + \bm{\omega}) \label{eq:q_star} \\
  \vec{y}^{\star} &=& \RB{\mat{F} + \frac{1}{MP}\vec{n}\bm{\omega}^T} \RB{\vec{q}^{\star}}^{-1}. \label{eq:y_star}
\end{IEEEeqnarray}
Substitute \eqref{eq:q_star} back to \eqref{eq:E}. Let $\mat{J}^{\star} = \mat{J}(\vec{q} = \vec{q}^{\star}, \vec{y} = \vec{y}^{\star})$ and $\mat{A} = \mat{J}^{\star} + \mat{I}$, we have
\begin{equation}
  \mat{A} =
  \left(
    \begin{array}{cc}
      \diag(\vec{q}^{\star}\circ\vec{y}^{\star})^{-1}\mat{F}^T & \diag\RB{\vec{q}^{\star}}\diag\RB{\vec{y}^{\star}}^{-1} \\
      \RB{\mat{F} + \frac{1}{MP}\vec{n}\bm{\omega}^T} \diag\RB{\vec{q}^{\star}}^{-2} & \mat{I} \\
    \end{array}
  \right).
\end{equation}
The matrix $\mat{A}$ is nonnegative and irreducible. Furthermore, we have
\begin{equation}
  \mat{A} \cdot \left(
            \begin{array}{c}
              \vec{q}^{\star} \\
              \vec{y}^{\star} \\
            \end{array}
          \right)
  + \left(
      \begin{array}{c}
        \diag(\vec{q}^{\star}\circ\vec{y})^{-1} \bm{\omega} \\
        \vec{0} \\
      \end{array}
    \right) = 2 \left(
            \begin{array}{c}
              \vec{q}^{\star} \\
              \vec{y}^{\star} \\
            \end{array}
          \right)
\end{equation}
according to \eqref{eq:q_star} and \eqref{eq:y_star}. Therefore,
\begin{equation}\label{eq:A}
  \mat{A} \cdot \left(
            \begin{array}{c}
              \vec{q}^{\star} \\
              \vec{y}^{\star} \\
            \end{array}
          \right) \lneqq
  2 \left(
            \begin{array}{c}
              \vec{q}^{\star} \\
              \vec{y}^{\star} \\
            \end{array}
          \right)
\end{equation}
and $\SB{\RB{\vec{q}^{\star}}^T, \RB{\vec{y}^{\star}}^T}^T \gneqq \vec{0}$. Because $\mat{A}$ is nonnegative and irreducible, \eqref{eq:A} implies its spectral radius $\rho(\mat{A}) < 2$ according to \cite[Theorem 1.11]{Berman1979nonnegative}. 



To ensure contraction mapping of the algorithm, the step \eqref{step:Aitken} must be invoked. Consider the update
\begin{equation}
  \vec{z}^{(m+1)} = \frac{1}{2} \vec{z}^{(m)} + \frac{1}{2} T(\vec{z}^{(m)}).
\end{equation}
Let $\vec{z}^{(m)} = \vec{z}^{\star} - \bm{\varepsilon}^{(m)}$ and $\vec{z}^{(m+1)} = \vec{z}^{\star} - \bm{\varepsilon}^{(m+1)}$, where $\vec{z}^{\star}$ is the optimum solution. Then
\begin{IEEEeqnarray}{rCl}
  \vec{z}^{\star} - \bm{\varepsilon}^{(m+1)} &=& \frac{1}{2} (\vec{z}^{\star} - \bm{\varepsilon}^{(m)}) + \frac{1}{2} T(\vec{z}^{\star} - \bm{\varepsilon}^{(m)}) \\
  &\approx& \frac{1}{2} (\vec{z}^{\star} - \bm{\varepsilon}^{(m)}) + \frac{1}{2} \RB{T(\vec{z}^{\star}) - \mat{J}^{\star} \cdot \bm{\varepsilon}^{(m)}} \\
  &=& \vec{z}^{\star} - \RB{\frac{1}{2} \mat{I} + \frac{1}{2} \mat{J}^{\star}} \bm{\varepsilon}^{(m)}.
\end{IEEEeqnarray}
Here we used $\vec{z}^{\star} = T(\vec{z}^{\star})$. Therefore,
\begin{equation}
  \bm{\varepsilon}^{(m+1)} \approx \RB{\frac{1}{2} \mat{I} + \frac{1}{2} \mat{J}^{\star}} \bm{\varepsilon}^{(m)}.
\end{equation}
Because $\rho(\mat{A}) < 2$, we have $\rho(\frac{1}{2} \mat{I} + \frac{1}{2} \mat{J}^{\star}) < 1$. This shows there exists neighborhood of the optimum solution, which ensures the mapping in Algorithm~\ref{alg:Alg1} to be a contraction mapping that satisfies the Lipschitz condition~\cite{granas2003fixed}. Therefore, Algorithm~\ref{alg:Alg1} converges to the optimum solution if the starting point is within this neighborhood.

\section*{Acknowledgment}
The authors would like to thank Prof. Chee Wei Tan and Mr. Xiangping Zhai for helpful discussions.

\bibliographystyle{IEEEtran}
\bibliography{\bibdir/header_short,\bibdir/bibliography}

\end{document}

%% file: Definitions.tex
\def\Real{\mathbb R}
\def\Complex{\mathbb C}

\newcommand{\norm}[1]{\left\Vert#1\right\Vert}
\newcommand{\abs}[1]{\left\vert#1\right\vert}

\newcommand{\E}{\mathrm{E}}

\newcommand{\EE}[1]{\ensuremath{\mathrm{E}\left\{#1\right\}}} 

\newcommand{\CN}{\ensuremath{\mathcal{CN}}}
\newcommand{\isCN}{\ensuremath{\sim\mathcal{CN}}}

\DeclareMathOperator{\tr}{tr}%
\DeclareMathOperator{\diag}{diag}%

\def\SINR{\textsf{SINR}}

\def\qt{\tilde{q}}
\def\as{\stackrel{\rm{a.s.}}{\longrightarrow}}

\newcommand{\RB}[1]{\left( #1 \right)}
\newcommand{\SB}[1]{\left[ #1 \right]}      
\newcommand{\CB}[1]{\left\{ #1 \right\}}    

\def\bm{\boldsymbol}
\def\mb{\mathbf}

\def\rm{\textrm}
\def\mc{\mathcal}

\newcommand{\mat}{\mb}      
\renewcommand{\vec}{\mb}    

%% file: JSAC_Revision_v2_2015_03_26.bbl
\begin{thebibliography}{10}
\providecommand{\url}[1]{#1}
\csname url@samestyle\endcsname
\providecommand{\newblock}{\relax}
\providecommand{\bibinfo}[2]{#2}
\providecommand{\BIBentrySTDinterwordspacing}{\spaceskip=0pt\relax}
\providecommand{\BIBentryALTinterwordstretchfactor}{4}
\providecommand{\BIBentryALTinterwordspacing}{\spaceskip=\fontdimen2\font plus
\BIBentryALTinterwordstretchfactor\fontdimen3\font minus
  \fontdimen4\font\relax}
\providecommand{\BIBforeignlanguage}[2]{{%
\expandafter\ifx\csname l@#1\endcsname\relax
\typeout{** WARNING: IEEEtran.bst: No hyphenation pattern has been}%
\typeout{** loaded for the language `#1'. Using the pattern for}%
\typeout{** the default language instead.}%
\else
\language=\csname l@#1\endcsname
\fi
#2}}
\providecommand{\BIBdecl}{\relax}
\BIBdecl

\bibitem{Andrews2014JSAC}
J.~G. Andrews, S.~Buzzi, W.~Choi, S.~Hanly, A.~Lozano, A.~C.~K. Soong, and
  J.~C. Zhang, ``What will {5G} be?'' \emph{IEEE J. Select. Areas Commun.},
  vol.~32, no.~6, pp. 1065--1082, Jun. 2014.

\bibitem{3GPPTR36913}
{3GPP TR 36.913}, ``Requirements for further advancements for {E-UTRA} ({LTE
  Advanced}),'' Nov. 2012, v11.0.0.

\bibitem{Chia2009Backhaul}
S.~Chia, M.~Gasparroni, and P.~Brick, ``The next challenge for cellular
  networks: Backhaul,'' \emph{IEEE Microw. Mag.}, vol.~10, no.~5, pp. 54--66,
  August 2009.

\bibitem{Yang2014twc}
Y.~Yang, T.~Q.~S. Quek, and L.~Duan, ``Backhaul-constrained small cell
  networks: Refunding and {QoS} provisioning,'' \emph{IEEE Trans. Wireless
  Commun.}, vol.~13, no.~9, pp. 5148--5161, Sep. 2014.

\bibitem{Hur2013mmWave}
S.~Hur, T.~Kim, D.~Love, J.~Krogmeier, T.~Thomas, and A.~Ghosh, ``Millimeter
  wave beamforming for wireless backhaul and access in small cell networks,''
  \emph{IEEE Trans. Commun.}, vol.~61, no.~10, pp. 4391--4403, Oct. 2013.

\bibitem{Lee2006wcnc}
S.~Lee, G.~Narlikar, M.~Pal, G.~Wilfong, and L.~Zhang, ``Admission control for
  multihop wireless backhaul networks with {QoS} support,'' in \emph{Proc. IEEE
  Wirel. Comm. and Netw. Conf.}, Las Vegas, NV, Apr. 2006.

\bibitem{Bojic2013advanced}
D.~Bojic, E.~Sasaki, N.~Cvijetic, T.~Wang, J.~Kuno, J.~Lessmann, S.~Schmid,
  H.~Ishii, and S.~Nakamura, ``Advanced wireless and optical technologies for
  small-cell mobile backhaul with dynamic software-defined management,''
  \emph{IEEE Commun. Mag.}, vol.~51, no.~9, pp. 86--93, Sep. 2013.

\bibitem{Ge2014FiveG}
X.~Ge, H.~Cheng, M.~Guizani, and T.~Han, ``{5G} wireless backhaul networks:
  challenges and research advances,'' \emph{IEEE Netw.}, vol.~28, no.~6, pp.
  6--11, Nov. 2014.

\bibitem{Viswanathan2006throughput}
H.~Viswanathan and S.~Mukherjee, ``Throughput-range tradeoff of wireless mesh
  backhaul networks,'' \emph{IEEE J. Select. Areas Commun.}, vol.~24, no.~3,
  pp. 593--602, Mar. 2006.

\bibitem{Zhao2012TWC}
J.~Zhao, T.~Q.~S. Quek, and Z.~Lei, ``Coordinated multipoint transmission with
  limited backhaul data transfer,'' \emph{IEEE Trans. Wireless Commun.},
  vol.~12, no.~6, pp. 2762--2775, Jun. 2013.

\bibitem{zhou2013uplink}
L.~Zhou and W.~Yu, ``Uplink multicell processing with limited backhaul via
  per-base-station successive interference cancellation,'' \emph{IEEE J.
  Select. Areas Commun.}, vol.~31, no.~10, pp. 1981--1993, Oct. 2013.

\bibitem{Park2013TSP}
S.-H. Park, O.~Simeone, O.~Sahin, and S.~Shamai, ``Joint precoding and
  multivariate backhaul compression for the downlink of cloud radio access
  networks,'' \emph{IEEE Trans. Signal Process.}, vol.~61, no.~22, pp.
  5646--5658, Nov. 2013.

\bibitem{Rao2013tvt}
J.~Rao and A.~Fapojuwo, ``On the tradeoff between spectral efficiency and
  energy efficiency of homogeneous cellular networks with outage constraint,''
  \emph{IEEE Trans. Veh. Technol.}, vol.~62, no.~4, pp. 1801--1814, May 2013.

\bibitem{Shi2014twc}
Y.~Shi, J.~Zhang, and K.~B. Letaief, ``Group sparse beamforming for green
  {Cloud-RAN},'' \emph{IEEE Trans. Wireless Commun.}, vol.~13, no.~5, pp.
  2809--2823, May 2014.

\bibitem{toelli2011decentralized}
A.~T\"olli, H.~Pennanen, and P.~Komulainen, ``Decentralized minimum power
  multi-cell beamforming with limited backhaul signaling,'' \emph{IEEE Trans.
  Wireless Commun.}, vol.~10, no.~2, pp. 570--580, Feb. 2011.

\bibitem{cai2012maxmin}
D.~W.~H. Cai, T.~Q.~S. Quek, C.~W. Tan, and S.~H. Low, ``Max-min {SINR}
  coordinated multipoint downlink transmission -- {Duality} and algorithms,''
  \emph{IEEE Trans. Signal Process.}, vol.~60, no.~10, pp. 5384--5395, Oct.
  2012.

\bibitem{zhai2014energy}
X.~Zhai, L.~Zheng, and C.~W. Tan, ``Energy-infeasibility tradeoff in cognitive
  radio networks: Price-driven spectrum access algorithms,'' \emph{IEEE J.
  Select. Areas Commun.}, vol.~32, no.~3, pp. 528--538, Mar. 2014.

\bibitem{Zhao2014TVT}
J.~Zhao, T.~Q.~S. Quek, and Z.~Lei, ``User admission and clustering for uplink
  multiuser wireless systems,'' \emph{IEEE Trans. Veh. Technol.}, vol.~64,
  no.~2, pp. 636--651, Feb. 2015.

\bibitem{ahmad2012coordinated}
T.~Ahmad, R.~Gohary, H.~Yanikomeroglu, S.~Al-Ahmadi, and G.~Boudreau,
  ``Coordinated port selection and beam steering optimization in a multi-cell
  distributed antenna system using semidefinite relaxation,'' \emph{IEEE Trans.
  Wireless Commun.}, vol.~11, no.~5, pp. 1861--1871, May 2012.

\bibitem{marzetta2010noncooperative}
T.~L. Marzetta, ``Noncooperative cellular wireless with unlimited numbers of
  base station antennas,'' \emph{IEEE Trans. Wireless Commun.}, vol.~9, no.~11,
  pp. 3590--3600, Sep. 2010.

\bibitem{rusek2013scaling}
F.~Rusek, D.~Persson, B.~K. Lau, E.~G. Larsson, T.~L. Marzetta, O.~Edfors, and
  F.~Tufvesson, ``Scaling up {MIMO}: Opportunities and challenges with very
  large arrays,'' \emph{IEEE Signal Processing Mag.}, vol.~30, no.~1, pp.
  40--60, Jan. 2013.

\bibitem{hoydis2013massive}
J.~Hoydis, S.~Ten~Brink, and M.~Debbah, ``Massive {MIMO} in the {UL/DL} of
  cellular networks: How many antennas do we need?'' \emph{IEEE J. Select.
  Areas Commun.}, vol.~31, no.~2, pp. 160--171, Feb. 2013.

\bibitem{fernandes2013inter}
F.~Fernandes, A.~E. Ashikhmin, and T.~L. Marzetta, ``Inter-cell interference in
  noncooperative {TDD} large scale antenna systems,'' \emph{IEEE J. Select.
  Areas Commun.}, vol.~31, no.~2, pp. 192--201, Feb. 2013.

\bibitem{Huang2013joint}
Y.~Huang, C.~W. Tan, and B.~Rao, ``Joint beamforming and power control in
  coordinated multicell: Max-min duality, effective network and large system
  transition,'' \emph{IEEE Trans. Wireless Commun.}, vol.~12, no.~6, pp.
  2730--2742, Jun. 2013.

\bibitem{matskani2008convex}
E.~Matskani, N.~Sidiropoulos, Z.~Luo, and L.~Tassiulas, ``Convex approximation
  techniques for joint multiuser downlink beamforming and admission control,''
  \emph{IEEE Trans. Wireless Commun.}, vol.~7, no.~7, pp. 2682--2693, Jul.
  2008.

\bibitem{cvx}
M.~Grant and S.~Boyd, ``{CVX}: Matlab software for disciplined convex
  programming, version 2.1,'' \url{http://cvxr.com/cvx}, Mar. 2014.

\bibitem{bengtsson2001optimal}
M.~Bengtsson and B.~Ottersten, ``Optimal and suboptimal transmit beamforming,''
  in \emph{Handbook of Antennas in Wireless Communications}, L.~C. Godara,
  Ed.\hskip 1em plus 0.5em minus 0.4em\relax Boca Raton, FL: CRC Press, 2001.

\bibitem{cai2011unified}
D.~W.~H. Cai, T.~Q.~S. Quek, and C.~W. Tan, ``A unified analysis of max-min
  weighted {SINR} for {MIMO} downlink system,'' \emph{IEEE Trans. Signal
  Process.}, vol.~59, no.~8, pp. 3850--3862, Aug. 2011.

\bibitem{zhang2012multi}
L.~Zhang, R.~Zhang, Y.-C. Liang, Y.~Xin, and H.~V. Poor, ``On {Gaussian MIMO
  BC-MAC} duality with multiple transmit covariance constraints,'' \emph{IEEE
  Trans. Inform. Theory}, vol.~58, no.~4, pp. 2064--2078, Apr. 2012.

\bibitem{dahrouj2010coordinated}
H.~Dahrouj and W.~Yu, ``Coordinated beamforming for the multicell multi-antenna
  wireless system,'' \emph{IEEE Trans. Wireless Commun.}, vol.~9, no.~5, pp.
  1748--1759, May 2010.

\bibitem{bai2009spectral}
Z.~D. Bai and J.~W. Silverstein, \emph{Spectral analysis of large dimensional
  random matrices}, 2nd~ed.\hskip 1em plus 0.5em minus 0.4em\relax Springer,
  2009.

\bibitem{couillet2011random}
R.~Couillet and M.~Debbah, \emph{Random matrix methods for wireless
  communications}.\hskip 1em plus 0.5em minus 0.4em\relax Cambridge University
  Press, 2011.

\bibitem{silverstein1995empirical}
J.~W. Silverstein and Z.~D. Bai, ``On the empirical distribution of eigenvalues
  of a class of large dimensional random matrices,'' \emph{Journal of
  Multivariate analysis}, vol.~54, no.~2, pp. 175--192, 1995.

\bibitem{couillet2011deterministic}
R.~Couillet, M.~Debbah, and J.~W. Silverstein, ``A deterministic equivalent for
  the analysis of correlated {MIMO} multiple access channels,'' \emph{IEEE
  Trans. Inform. Theory}, vol.~57, no.~6, pp. 3493--3514, Jun. 2011.

\bibitem{Wagner2012LargeSystem}
S.~Wagner, R.~Couillet, M.~Debbah, and D.~T. Slock, ``Large system analysis of
  linear precoding in correlated {MISO} broadcast channels under limited
  feedback,'' \emph{IEEE Trans. Inform. Theory}, vol.~58, no.~7, pp.
  4509--4537, Jul. 2012.

\bibitem{bai2007signal}
Z.~D. Bai and J.~W. Silverstein, ``On the signal-to-interference ratio of
  {CDMA} systems in wireless communications,'' \emph{The Annals of Applied
  Probability}, vol.~17, no.~1, pp. 81--101, 2007.

\bibitem{feyzmahdavian2012contractive}
H.~R. Feyzmahdavian, M.~Johansson, and T.~Charalambous, ``Contractive
  interference functions and rates of convergence of distributed power control
  laws,'' \emph{IEEE Trans. Wireless Commun.}, vol.~11, no.~12, pp. 4494--4502,
  Dec. 2012.

\bibitem{schubert04solution}
M.~Schubert and H.~Boche, ``Solution of the multiuser downlink beamforming
  problem with individual {SINR} constraints,'' \emph{IEEE Trans. Veh.
  Technol.}, vol.~53, no.~1, pp. 18--28, 2004.

\bibitem{3GPPTS36814}
{3GPP TS 36.814}, ``Evolved universal terrestrial radio access ({E-UTRA});
  further advancements for {E-UTRA} physical layer aspects,'' Mar. 2010,
  v9.0.0.

\bibitem{stridh2006system}
R.~Stridh, M.~Bengtsson, and B.~Ottersten, ``System evaluation of optimal
  downlink beamforming with congestion control in wireless communication,''
  \emph{IEEE Trans. Wireless Commun.}, vol.~5, no.~4, pp. 743--751, Apr. 2006.

\bibitem{yoo2006optimality}
T.~Yoo and A.~Goldsmith, ``On the optimality of multiantenna broadcast
  scheduling using zero-forcing beamforming,'' \emph{IEEE J. Select. Areas
  Commun.}, vol.~24, no.~3, pp. 528--541, Mar. 2006.

\bibitem{Boche2005dualityBook}
H.~Boche and M.~Schubert, ``Duality theory for uplink downlink multiuser
  beamforming,'' in \emph{Smart Antennas -- State of the Art}.\hskip 1em plus
  0.5em minus 0.4em\relax Hindawi Publishing Corporation, 2005, ch.~27, pp.
  545--576.

\bibitem{van2002optimum}
H.~Van~Trees, \emph{Optimum array processing}.\hskip 1em plus 0.5em minus
  0.4em\relax John Wiley \& Sons, 2002.

\bibitem{Boyd04ConvexOptimization}
S.~Boyd and L.~Vandenberghe, \emph{Convex Optimization}.\hskip 1em plus 0.5em
  minus 0.4em\relax Cambridge University Press, 2004.

\bibitem{Berman1979nonnegative}
A.~Berman and R.~J. Plemmons, \emph{Nonnegative Matrices in the Mathematical
  Sciences}.\hskip 1em plus 0.5em minus 0.4em\relax New York: Academic Press,
  1979.

\bibitem{granas2003fixed}
A.~Granas and J.~Dugundji, \emph{Fixed point theory}.\hskip 1em plus 0.5em
  minus 0.4em\relax Springer, 2003.

\end{thebibliography}
